# Translation-Invariant Noncommutative Gauge Theories, Matrix Modeling and Noncommutative Geometry

Amir Abbass Varshovi

amirabbassv@ipm.ir

*School of Physics, Institute for Research in Fundamental Science (IPM)*

*P. O. Box 19395-5531, Tehran, IRAN*

*Department of Physics, Sharif University of Technology*

*P. O. Box 11365-9161, Tehran, IRAN*

**Abstract;** A matrix modeling formulation for translation-invariant noncommutative gauge theories is given in the setting of differential graded algebras and quantum groups. Translation-invariant products are discussed in the setting of $\alpha$-cohomology and it is shown that loop calculations are entirely determined by $\alpha$-cohomology class of star product in all orders. Noncommutative version of geometric quantization and (anti-) BRST transformations is worked out which leads to a noncommutative description of consistent anomalies and Schwinger terms.

## 1. Introduction

Using noncommutative structures for space-time coordinates at very small length scales, was suggested in the early years of quantum mechanics by its founding fathers to introduce an effective ultraviolet cutoff and to give an appropriate setting to describe the small scale structures of the universe [1, 2]. The first mathematical formulation of noncommutative coordinates was worked out by Snyder [3, 4]. This formulation was strongly motivated by the need to control the divergences of quantum electrodynamics in its very beginning formalisms.

But the success of renormalization program of field theories made the suggestion of noncommutative coordinates, to be largely ignored in its time. However the idea of noncommutative space-time revived after considering the quantization of gravity where it became clear that the space-time should be quantized in some way to make up a "pointless geometry". Indeed noncommutativity of coordinate functions asserts that they cannot be diagonalized simultaneously and thus the underlying point-wise concept of space disappears. It was von Neumann who first used the term of pointless geometry for his attempts to rigorously describe a quantum space [5]. Indeed failing the meaning of points in phase space of quantum mechanics due to the Heisenberg uncertainty principle, motivated him to look after a pointless geometry. Eventually the idea of pointless geometry led to the theory of von Neumann algebras as a foundation of noncommutative geometry.

More precisely, noncommutative geometry is a theory based on the non-commutative version of the celebrated Gelfand-Neimark theorem [6] which identifies the category of locally compact Hausdorff spaces and commutative unital $C^*$-algebras in a contravariant manner [7, 8]. More than topological point of views, the idea of noncommutative geometry was also generalized to differential structures. This generalization led to the concept of quantized calculus in the setting of differential graded algebras (DGA) [7-11] and differential calculus on quantum groups [12-14].

Quantum groups, as an independent development of noncommutative geometry, were first introduced by Drinfeld [15, 16] to provide solutions of the quantum Yang-Baxter equation in the setting of Hopf algebras which had been found by Hopf in his elaborations in algebraic topology. In fact, Drinfeld constructed a family of quantum groups, so called quantum doubles, as a mathematical formulation of the method due to Faddeev, Reshetikhin and Takhtadjian, in studying quantum Yang-Baxter equation and quantum inverse scattering. More precisely quantum groups are an exciting generalization of ordinary Lie groups in the sense that they have a rich mathematical structure and numerous roles in situations where ordinary Lie groups are not adequate [17-20].

The collection of noncommutative geometry and quantum groups is mostly referred to as "quantum geometry". Therefore in quantum geometry the classical intuitions of smooth manifolds and Lie groups were replaced by algebraic concepts. Although it was well-known that to do geometry on Riemannian manifolds and more specially on Lie groups, it is often convenient to work with the algebra of functions, but it was the idea of quantum geometry that even when this algebra is deformed or made to a noncommutative one, one can continue to do geometry, provided all our constructions are referred to algebras rather than to any underlying manifold which need no longer exist.

Noncommutative geometry has provided a set of enormous applications in quantum physics. The most sophisticated manifestation of noncommutative geometry in quantum physics is to describe the Standard Model in the setting of noncommutative Riemannian manifolds or spectral triples [21]. But the most popular application of noncommutative geometry in quantum physics is the studying of noncommutative quantum field theories. Indeed, the successes of noncommutative geometry in the realm of quantum physics led to a revival of Snyder's idea for noncommutative space-time at Planck scale [22, 23] and consequently led to noncommutative field theories. But it is not the whole story, since the idea of noncommutative field theories owes most of its appearance to developments of string theory where more evidences for space-time noncommutativity came from [24-27].

Noncommutative field theories have first studied in a naïve approach, using the Groenewold-Moyal product [28, 29] instead of the ordinary one. Developing Feynman rules [30] and considering the perturbation theory [31], revealed a serious problem in renormalization program of noncommutative scalar field theories, so called UV/IR mixing. UV/IR mixing is a pathological effect which plagues the theory by reflecting UV divergences in new IR singularities. Soon after it was shown [32-35] that noncommutative gauge theories are also suffered by UV/IR mixing. Although this pathology was cured for scalar field theories in translation-variant [36, 37] and non-local [38] formulations, but the problem is still unsolved for noncommutative gauge theories [39].

But as mentioned above, since string theory may be defined as an appropriate modification of classical general relativity, the need of noncommutative space-time in string theory is actually more apparent than in quantum field theory [40]. Despite of this fact, there has remained a gap in understanding of the role of noncommutative geometry in string theory [5]. Indeed, there is no natural dynamical origin for the occurrence of noncommutative field theories. This point may be explained by open string degrees of freedom known as D-branes [41], which are fixed hypersurfaces in space-time onto which the endpoints of strings can attach. It was also realized that the low-energy effective field theory of D-branes has configuration space which is described in non-commuting matrix valued space-time coordinate fields [42]. This has led to two suggestions for M theory as a rigorous generalization of superstring theory in which strings are on equal footing with their multidimensional analogues, branes. First of all it was conjectured [43] that M theory can be defined as a matrix quantum mechanics obtained from 10-dimensional supersymmetric Yang-Mills theory by means of reduction to $0 + 1$-dimensional theory where the size of the matrix tends to infinity. This matrix modeling of M theory is often referred to as BFSS matrix model which stands for its founders. The other matrix model [44], inspired by BFSS model, was obtained soon after by dimensionally reducing supersymmetric Yang-Mills theory to a point. This matrix modeling proposed a matrix base formulation for type IIB superstring theory in agreement with Green-Schwarz action [45] in the Schild gauge [46]. This matrix modeling of superstring theory is usually known as IKKT model. Then after it was shown [47] that the methods of noncommutative geometry can be used to analyze and formulate the BFSS and IKKT matrix models. Moreover, it was also shown that the Eguchi-Kawai reduction of IKKT model leads to BFSS model.

The idea of matrix modeling also affected noncommutative field theories [5] as was expected for their string theoretical backgrounds. Thus, noncommutative field theories were regulated and studied by means of matrix models [48-50]. This matrix regularization yield field theories on the fuzzy torus or the fuzzy sphere and showed an intimate relation between field theories on the noncommutative torus and the lattice regularization of noncommutative field theories [51-54]. These investigations led to a different interpretation of the origin of UV/IR mixing in noncommutative gauge models [55-59] by considering matrix model of Yang-Mills with;

$$S_{YM} = -Tr\{[X^a, X^b][X^c, X^d]\}\eta_{ac}\eta_{bd},$$

(1)

$1 \leq a, b, c, d \leq D$, for $\eta_{ab}$, the metric of some $D$-dimensional manifold and for Hermitian matrices $X^a \in M_N(\mathbb{C})$ which act on a Hilbert space $\mathcal{H}$. In the simplest case $X^a$s represent "generalized coordinates" and in the semi-classical limit $X \sim x$ one can interpret them as defining the embedding of a $2n$-dimensional submanifold $\mathcal{M}^{2n} \subseteq \mathbb{R}^D$ equipped with a non-trivial metric

$$g_{\mu\nu} = \partial_\mu x^a \partial_\nu x^b \eta_{ab},$$

(2)

via pull-back of $\eta_{ab}$ [39]. This submanifold could then be considered as the 4-dimansional noncommutative space-time with the Poisson structure

$$\theta^{\mu\nu} \sim -i[X^\mu, X^\nu].$$

(3)

It can be shown that UV/IR mixing can be considered as a good candidate to reconcile the apparent nonrenormalizability of gravity with nice renormalization behavior of Yang-Mills theories in gravitational descriptions of matrix model (1) [60].

But as mentioned above curing UV/IR mixing in noncommutative field theories leads to more serious problems of breaking the translation-invariance [36, 37] or the locality [38]. Breaking the translation-invariance leaves a quantum theory meaningless by losing the energy-momentum conservation law. This provided motivations for studying a large family of generalized Groenewold-Moyal star product which do not depend on coordinate functions. This family of star products is commonly referred to as "translation-invariant products" [61, 62] since they keep the translation-invariance of the theory when they are used instead of the ordinary product. Such restricted class of deformation quantization [63-65] of the algebra of functions is intimately related to the chosen coordinate system on the underlying space-time manifold. Although Lorentz transformations of coordinate systems preserve translation-invariant star products, but it is known that noncommutative field theories with translation-invariant star products cannot be relativistic [66-68]. But while the noncommutativity of space-time ruins the causality [66, 67], the unitarity [68] and consequently the Lorentz invariance of quantum field theories, the studying of noncommutative field theories is still of interest for their stringy spirit, where no acausal behavior is seen and the unitarity is preserved.

Translation-invariant quantum field theories with translation-invariant non-commutative products are usually referred to as "translation-invariant non-commutative quantum field theories". It is known [61] that, the quantum behaviors of translation-invariant noncommutative scalar and gauge field theories at one-loop corrections, are entirely described by commutators of coordinate functions. Moreover, the associativity condition of translation-invariant products leads to a cohomology theory, so called $\alpha$-cohomology, which is intimately related to Hochschild cohomology [7] and classifies the star products up to a family of commutative products due to coboundaries [61].

On the other hand, as well as noncommutative geometry, quantum groups provide a number of extremely large developments of quantum physics such as discrete gauge theories, topological quantum field theories, lattice gauge theories, $2 + 1$-dimensional topological Aharonov-Bohm effect [69, 70], the theory of quantum statistics [71], the theory of anyons [72], the theory of topological quantum computation and the theory of fractional quantum Hall effect [73, 74]. Moreover, although the theory of quantum groups was a spectacular achievement of theoretical physics, but it produced a collection of rigorous developments in a large breadth of remote domains in mathematics such as the theories of von Neumann algebras, representation of semisimple Lie algebras, Fusion theory, low dimensional topology and the theory of knots [18, 19, 20, 75, 76].

In this article a new matrix modeling is proposed for translation-invariant noncommutative gauge (TNG) theories in the setting of DGAs and quantum groups. This formulation enlarges the class of gauge theories for enormous variety of quantum groups. Indeed, the large extent of deforming methods of Lie groups and Lie algebras in the context of quantum groups leads to an extremely larger class of translation-invariant noncommutative quantum field theories with gauge theoretical spirits. More precisely, the crucial role of principal bundles in the definition of gauge theories is removed in this matricial formulation of gauge theories. Therefore, these matrix modeled gauge (MMG) theories can be considered as a generalized version for translation-invariant gauge theories.

In this article it is also shown that for any TNG theory, there exists an appropriate family of MMG theories which can describe the theory in large $N$ limits. Moreover, MMG theories are quantized by geometric quantization approach and this leads to a noncommutative version of (anti-) BRST transformations. Consequently, this matricial formulation of TNG theories leads to a noncommutative counterpart of descent equations and consistent anomalies. On the other hand, the theory of $\alpha$-cohomology is studied and an algebraic Hodge theorem is worked out for $\alpha$-cohomology groups. This leads to a fascinating classification of translation-invariant products.

In section 2, Matrix Modeling of Gauge theories, the basic idea of MMG theories is worked out by technical elaborations and mathematical details. Initially the simplest formulation of MMG theories, called primitive MMG theories, is defined in the setting of DGA's and quantum groups. It is seen that primitive MMG theories leads to commutative Yang-Mills theories with extremely more degrees of freedom which has no desirable feature. To overcome these difficulties one should go further and define MMG theories by the conjecture of matricial quantization map (MQM). Finally it is shown that any translation-invariant noncommutative Yang-Mills theory can be described by a MMG theory.

In section 3, Matrix Modeling and Translation-Invariant Noncommutative Gauge Theories, it is shown that the conjecture of MQM comes true by large $N$ limit of a family of matricial formulations and consequently an equivalency correlation is proven for MMG theories and TNG theories.

In section 4, Translation-Invariant Deformation Quantization, translation-invariant star products are discussed in the setting of $\alpha$-cohomology. It is strictly shown that the second $\alpha$-cohomology group classifies the translation-invariant star products modulo commutative ones. Also an algebraic version of Hodge theorem is proven for second $\alpha$-cohomology group. This leads to a unique representing element for each $\alpha$-cohomology class called harmonic form. Using the concept of harmonic forms, it is shown that the quantum corrections of translation-invariant noncommutative field theories are entirely described by $\alpha$-cohomology classes in all orders.

In section 5, Quantization and Consistent Anomalies of Matrix Modeled Gauge Theories, MMG theories are quantized by the approach of geometric quantization. A noncommutative version for (anti-) BRST transformations and eventually for descent equations is worked out in the setting of matrix modeling. This consequently leads to a noncommutative counterpart of consistent anomalies and consistent Schwinger terms for TNG theories. MMG theories are also quantized in the path integral formalism due to modified partition functions method. The consistent anomalies and Schwinger terms extracted by geometric quantization method are confirmed by the results of modified partition functions.

In section 6, Summery and Conclusions, the article achievements are listed in brief.

In section 7, three appendices are attached each of which includes a detailed proof of a stated theorem in the article which may be remote from the body of discussions.

# 2. Matrix Modeling of Gauge Theories

In this section, matrix modeled gauge (MMG) theories are introduced and their properties are discussed in a very general framework. To this end for all this section it is supposed that the space-time, $M$, takes the form of $\mathbb{R}^{d_1} \times \mathbb{T}^{d_2}$, $d_1 \geq 1, D = d_1 + d_2$, endowed with the Minkovskian metric. It is also assumed that time is a non-cyclic coordinate, $x^0(M) = \mathbb{R}$. As a more general setting, one can extend all contents of this section to any m-dimensional spin manifold equipped with a semi-Riemannian metric with some appropriate technicalities. Conventionally $S(M)$ is used for the spin bundle of $M$ and its fiber, the standard spinor space is shown by $S$. More precisely the spinor bundle is represented by; $S \hookrightarrow S(M) \twoheadrightarrow M$. One also needs a connection over $S(M)$ with covariant derivative $\nabla_{S(M)}$. For simplicity it is supposed that $\nabla_{S(M)}$ defines a flat connection over $S(M)$. Basically to build an ordinary gauge theory over $M$, one needs a principal $G$-bundle, say $G \hookrightarrow P \twoheadrightarrow M$, a connection over it and a unitary irreducible representation of the semi-simple gauge group $G$ over the representing complex vector space $V$. This representation induces a vector bundle over $M$, shown by; $V \hookrightarrow E := P \times_G V \twoheadrightarrow M$. On the other hand, the connection of $G \hookrightarrow P \twoheadrightarrow M$ defines naturally a connection over $V \hookrightarrow E \twoheadrightarrow M$ with covariant derivative $\nabla_E$. A $G$-gauge theory over $M$ is defined over the tensor product vector bundle $S(M) \otimes E \twoheadrightarrow M$ with the induced covariant derivative; $\nabla_{S(M)} \otimes 1_E + 1_{S(M)} \otimes \nabla_E$. A $G$-colored spinor is a section of this vector bundle and a $G$-gauge theory is defined by a functional of these $G$-colored spinors and their covariant derivatives. Generally for the base space $M$, one can always define the covariant derivative $\nabla_E$ by $d + A$ for some $\mathfrak{g}$-valued 1-form $A$, where $\mathfrak{g}$ is the Lie algebra of $G$. Usually these operator valued 1-forms are called gauge fields. In fact, any ordinary gauge theory can also be defined locally by such operator valued 1-forms. It is enough to define an irreducible Hermitian representation of $\mathfrak{g}$ over the inner product space $V$ and to use the local coordinate systems as trivializations of the relevant vector bundle. Since the gauge group elements can be considered as a finite product of the exponentials of elements of $\mathfrak{g}$, the action of gauge group would be well-defined. In the following the same idea is used to define an $N \times N$-matrix modeled gauge theory.

## 2.1. Primitive Matrix Modeled Gauge Theories

Before defining a MMG theory over $M$, one should start with definition of primitive MMG theories. For the beginning, the primitive $N \times N$-matrix modeled $U(1)$-gauge theory is discussed.

Consider the trivial vector bundle $S(M)_N := S(M) \otimes M_N(\mathbb{C}) \twoheadrightarrow M$ with $M_N(\mathbb{C})$ the space of $N \times N$ complex matrices and let $H$ be the tensor algebra of $C^\infty(M, M_N(\mathbb{C})) = C^\infty(M) \otimes M_N(\mathbb{C})$ over $\mathbb{C}$. Here $H$ is considered to be a Hopf algebra with the ordinary unit, the product $\otimes$, the counit $\varepsilon(a) = 0$, the co-product $\Delta(a) = a \otimes 1 + 1 \otimes a$ and the antipode $S(a) = -a$, $a \in C^\infty(M) \otimes M_N(\mathbb{C})$ [18, 19]. $H$ can also be equipped with more structures, like involution $*$, which is given by $a^* = a^\dagger$, $a \in C^\infty(M) \otimes M_N(\mathbb{C})$ and can be extended to all of $H$ by considering that $*$ is an algebra anti-morphism. Next one may need a differential graded algebra (DGA) [9-12] $(\Omega = \oplus_d \Omega^d, d)$ over $\Omega^0 = H$. To this end, set a fixed coordinate system over $M$, $\{x^i\}_{i=0}^{D-1}$, with $x^i = x^i_{\mathbb{R}^{d_1}} : M \twoheadrightarrow \mathbb{R}$, for $0 \leq i < d_1$, and $x^i = x^{i-d_1}_{\mathbb{T}^{d_2}} : M \hookrightarrow \mathbb{R}$, for $d_1 \leq i < D$. Now consider the split Poincare group, $\mathcal{SP}(M)$, including of ordinary translations together with rotations and boosts of which do not mix $x^i_{\mathbb{R}^{d_1}}$ and $x^i_{\mathbb{T}^{d_2}}$. It can be easily checked that $\mathcal{SP}(M)$ is a group. The coordinate system $\{x'^i\}_{i=0}^{D-1}$ over $M$, is called $\mathcal{SP}$-equivalent to $\{x^i\}_{i=0}^{D-1}$ if there exists $\Lambda \in \mathcal{SP}(M)$ such that $x'^i = x^i \circ \Lambda$. This defines an equivalence class of coordinate systems, say $\mathcal{SP}$-class. A field theory is $\mathcal{SP}$-invariant if its action is invariant under $\mathcal{SP}(M)$ transformations. In fact, from now on the Poincare invariance is replaced by $\mathcal{SP}$-invariance in our formulation. Indeed there exists a natural right action of $\mathcal{SP}(M)$ on $H$ by pulling back the elements of $H$ through the elements of $\mathcal{SP}(M)$. Actually we are interested in the action of translation subgroup $\mathcal{T}(M) \subseteq \mathcal{SP}(M)$. This leads to dynamical system $(H, \mathcal{T}(M), \tau)$ with $\tau$ the pull-back representation. As it is shown in the following this dynamical system helps one to construct a DGA over $H$ in a consistent manner.

The differential operator, d, is given over $H$ with $d1 := 0$ and $da := \partial_\mu a \, dx^\mu$ for any $a \in C^\infty(M) \otimes M_N(\mathbb{C})$. It is seen that d is uniquely defined on $\mathcal{SP}$-class charts and thus is well-defined. Indeed, d can be expressed as the representation of Lie $\mathcal{T}(M)$ in the Lie algebra of derivations of $H$ given by; $d_X a = \frac{d}{dt}\Big|_{t=0} a \circ X_t$, for $a \in H$ and $X_t$ the 1-parameter group for $X \in \text{Lie } \mathcal{T}(M)$. Moreover, d can be uniquely defined over the whole of $\Omega$ satisfying the graded Leibnitz rule and the nilpotency condition $d^2 = 0$. Specially, d is an anti-derivative with degree 1 given by

$$d\left(\sum a_{\mu_1 \ldots \mu_d} dx^{\mu_1} \wedge \ldots \wedge dx^{\mu_d}\right) = \sum da_{\mu_1 \ldots \mu_d} \wedge dx^{\mu_1} \wedge \ldots \wedge dx^{\mu_d}$$

(4)

for $a_{\mu_1 \ldots \mu_d} \in H$.

From (4) it can be easily checked that $d^2 = 0$ and for any two homogeneous elements $\omega_1, \omega_2 \in \Omega$, one finds that; $d(\omega_1 \omega_2) = d\omega_1 \omega_2 + (-)^{|\omega_1|} \omega_1 d\omega_2$.

Moreover, it is seen that in general $\omega_1 \omega_2 \neq (-)^{|\omega_1||\omega_2|} \omega_2 \omega_1$. In fact, $\Omega^d$ is precisely equal to $H \otimes \Omega^d_{\text{deR}/\mathcal{SP}}(M)$ with $\Omega^d_{\text{deR}/\mathcal{SP}}(M) = \Lambda^d(\text{Lie } \mathcal{T}(M))^*$. Indeed $\Omega^d_{\text{deR}/\mathcal{SP}}(M)$ is the chain of deRham $d$-forms over $M$ which are expanded by terms of $dx^{\mu_1} \wedge \ldots \wedge dx^{\mu_d}$, $\mu_1 < \cdots < \mu_d$ over $\mathbb{C}$. Actually, $\Omega^d$ is a free $H$-bimodule generated with elements $dx^{\mu_1} \wedge \ldots \wedge dx^{\mu_d}$, $\mu_1 < \cdots < \mu_d$. Also it can be easily seen that $\Omega$ can be equipped with a natural Hodge star structure by;

$$\star \left(\sum a_{\mu_1 \ldots \mu_d} dx^{\mu_1} \wedge \ldots \wedge dx^{\mu_d}\right) := \sum a_{\mu_1 \ldots \mu_d} \star (dx^{\mu_1} \wedge \ldots \wedge dx^{\mu_d}),$$

(5)

for any $\sum a_{\mu_1 \ldots \mu_d} dx^{\mu_1} \wedge \ldots \wedge dx^{\mu_d} \in \Omega$.

Furthermore, one may need to equip $\Omega$ with a right comodule structure by $\varphi \in \text{Hom}_\mathbb{C}(\Omega, \Omega \otimes H)$ with

$$\varphi\left(\sum a_{\mu_1\ldots\mu_d}dx^{\mu_1}\wedge\ldots\wedge dx^{\mu_d}\right) := \sum (a_{\mu_1\ldots\mu_d})_{(1)}dx^{\mu_1}\wedge\ldots\wedge dx^{\mu_d} \otimes (a_{\mu_1\ldots\mu_d})_{(2)},$$

(6)

for $a_{\mu_1\ldots\mu_d} \in H$ and for the standard Sweedler notation of $\Delta(a) = \sum a_{(1)} \otimes a_{(2)}$, $a \in H$. Obviously, the comodule structure (6) is compatible with unit and multiplication of $\Omega$. Therefore, (6) makes $\Omega$ to be a right $H$-comodule algebra. In other words, $\Omega$ is a right quantum space over $H$ [20].

Here $\Omega$ does not need to be a right covariant differential calculus [20]. Indeed $\varphi \circ d \neq (d \otimes id_H) \circ \varphi$ generally. Actually it is enough to know that $\Omega$ is an algebra in the category of right $H$-comodules and consequently $\Omega$ is an Yetter-Drinfeld module [77]. Conventionally, the notation of $\sum \omega_{(0)} \otimes \omega_{(1)}$ is used for $\varphi(\omega)$, $\omega \in \Omega$, according to [20].

Now consider the elements of $\Omega^0 = H$ as the space of infinitesimal gauge transformations which act on the elements of $\Gamma(S(M)_N)$ in various modes;

- Type I) The ordinary action;

$$\sum a_{i_1} \otimes \ldots \otimes a_{i_k} \triangleright_I \psi := \sum a_{i_1}\ldots a_{i_k} \psi.$$

(7)

- Type II) The inverse action;

$$a \triangleright_{II} \psi := \sum \psi \triangleleft S(a).$$

(8)

- Type III) The adjoint action;

$$a \triangleright_{III} \psi := ad(a)(\psi) = \sum a_{(1)} \triangleright_I \psi \triangleleft S(a_{(2)}).$$

(9)

with $\psi \in \Gamma(S(M)_N)$ and $a, \sum a_{i_1} \otimes \ldots \otimes a_{i_k} \in H$. The right action $\psi \triangleleft h, h \in H$ in (8) and (9) is defined similar to (7) but from the right hand side. Moreover, the elements of $\Omega^d$, with $d \geq 1$, act similarly on the elements of $\Gamma(S(M)_N)$ in three given types (7)-(9). This leads to matricial spinor valued $d$-forms. On the other hand, one can consider the derivative operator d to act on matricial spinors in the natural way which sends the elements of $\Gamma(S(M)_N)$ to the elements of $\Gamma(S(M)_N) \otimes \Omega^1_{\text{deR}/\mathcal{SP}}(M)$.

Although for any three types of actions (7)-(9), $\Gamma(S(M)_N)$ is not a projective module over $H$, but there can be found a connection structure and eventually a curvature 2-form over it according to [78]. As it was stated above, setting $\tau_f = \delta f$, the pull back through $f \in \mathcal{T}(M)$, makes $(H, \mathcal{T}(M), \tau)$ to be a dynamical system which induces the differential structure d over $H$. Fortunately this differential structure naturally leads to a flat connection and consequently to an algebraic covariant derivation on $\Gamma(S(M)_N)$ [79] similarly denoted by d. Moreover, any linear map $\nabla = \text{d} + iA$, for any $A \in \Omega^1$ and all types of actions (7)-(9), sends the elements of $\Gamma(S(M)_N)$ to the set of matricial spinor valued 1-forms, $\Gamma(S(M)_N) \otimes (\text{Lie } \mathcal{T}(M))^* = \Gamma(S(M)_N) \otimes \Omega^1_{\text{deR}/\mathcal{SP}}(M)$. In fact, $\nabla$ defines a connection over $\Gamma(S(M)_N)$. Therefore, one may consider the elements of $\Omega^1$ as the gauge fields. Moreover, the covariant derivative can be extended to $\Gamma(S(M)_N) \otimes \Omega^d_{\text{deR}/\mathcal{SP}}(M)$ for $d \geq 1$ naturally. Thus, it can be directly checked that $\hat{R} = i\text{d}A - A^2 \in \Omega^2$ is the curvature of connection $\nabla = \text{d} + iA$ [78] for all types of actions (7)-(9). On the other hand, the Dirac operator $\mathcal{D}$ can be defined analogously by $\mathcal{D} := \gamma^\mu \nabla_\mu$, for the Clifford algebra generators $\gamma^\mu$, $\mu = 1, \ldots, D$.

To define the action of a field theory one obviously needs an integration structure. Indeed, there is a natural integration structure over $(\Omega, \text{d})$ which makes $(H, \Omega, \int)$ to be a $D$-dimensional cycle over $H$ according to [9];

$$\int \sum a_{i_1} \otimes \ldots \otimes a_{i_k} \, \text{d}x^1 \wedge \ldots \wedge \text{d}x^D := \sum \int_M tr\{a_{i_1} \ldots a_{i_k}\},$$

(10)

for any $\sum a_{i_1} \otimes \ldots \otimes a_{i_k} \in H$. It can be easily seen that for any set of elements $\omega, \omega_1, \omega_2 \in \Omega$, one finds that $\int d\omega = 0$ and $\int \omega_1 \omega_2 = (-)^{|\omega_1||\omega_2|} \int \omega_2 \omega_1$. Thus, $\int$ gives hand a well-defined integration structure over $\Omega$.

Now it is the time to give the explicit form of the Lagrangian for a primitive $N \times N$-matrix modeled $U(1)$-gauge theory. Here one should assume that the Clifford algebra elements $\gamma^\mu$ are replaced with $\gamma^\mu \otimes \star (dx^\mu)$ (without summing over $\mu$). Thus the primitive form of the Lagrangian density of matter is defined by

$$\mathcal{L}_{Matter} := \langle \psi | i\mathcal{D}\psi \rangle := \bar{\psi}\, i\mathcal{D}\psi = \psi^\dagger \gamma^0 i\mathcal{D}\psi\,,$$

(11)

where $\langle\,|\,\rangle$ is a sesquilinear map, $\langle\,|\,\rangle : S(M)_N \otimes S(M)_N \to \Omega$, induced naturally by the standard Hermitian inner product over $S(M)$. It is seen that the Lagrangian (11) is a matrix valued functional and is an element of $\Omega^D$ which can be integrated by (10) to give hand the action of matter fields. Indeed, to have a real valued action one should refine the definition of gauge fields and restrict them to Hermitian elements of $\Omega^1$, $A^* = A$. On the other hand, note that any type of actions (7)-(9) leads to a particularly distinct theory as will be cleared in the following.

Also for even dimensional cases the Maxwell Lagrangian densities are given analogously in terms of the curvature, $\hat{R}$, and its Hodge dual $\star \hat{R}$. For example in 4-dimensional case one can set;

$$\mathcal{L}_{Maxwell} = \hat{R} \star \hat{R}\,.$$

(12)

Thus, the Maxwell Lagrangian density also lies in $\Omega^D$ and therefore by (10) the complete action is;

$$S = S_{Maxwell} + S_{Matter} := \int \mathcal{L}_{Maxwell} + \int \mathcal{L}_{Matter}\,.$$

(13)

To check the gauge invariance of this action one needs to introduce the infinitesimal gauge transformations of matter and gauge fields. The gauge transformations of matter fields are defined by actions of $H$ on $\Gamma(S(M)_N)$, with all three types (7)-(9). Thus it remains to define a right action of $H$ on $\Omega^1$. Set

$$A \cdot \alpha = id\alpha + [A, \alpha]_{\text{Act.}},$$

(14)

for $\alpha \in H$ and $A \in \Omega^1$. Here the *action commutator*, $[\,,\,]_{\text{Act.}}$, is computed with considering the relevant action types of the elements. Indeed, the action types of $A$ and $\alpha$ may be different and the action commutator respects this fact. For example the ordinary (type I) and the inverse (type II) acting elements always commute for $[\,,\,]_{\text{Act.}}$. On the other hand, according to (14), the infinitesimal gauge transformation of the curvature is given by;

$$\hat{R} \cdot \alpha = [\alpha, \hat{R}]_{\text{Act.}},$$

(15)

which keeps the Maxwell action, $S_{Maxwell}$, invariant. The variation of matter action under infinitesimal gauge transformation $\alpha$ is defined by;

$$\delta S_{Matter} := \int \langle \delta_\alpha \psi | i\mathcal{D}\psi \rangle + \int \langle \psi | i\mathcal{D} \delta_\alpha \psi \rangle + \int \langle \psi | i\delta_\alpha(\mathcal{D})\psi \rangle,$$

(16)

where $\delta_\alpha \psi = \alpha \triangleright_i \psi$ for $i = \text{I, II, III}$, and $\delta_\alpha(\mathcal{D}) = iA \cdot \alpha$ for $\nabla = d + iA$. Actually since $(a \triangleright_\text{I} \psi)^\dagger = \psi^\dagger \triangleleft a^*$ (resp. $(\psi \triangleleft a)^\dagger = a^* \triangleright_\text{I} \psi^\dagger$), $a \in H$, it is clear that to have $\delta S_{Matter} = 0$, $\alpha$ must be an anti-Hermitian element of $\Omega^0 = H$, i.e. $\alpha^* = -\alpha$. Thus, the definition of infinitesimal gauge transformations should be modified to include only the anti-Hermitian elements of $\Omega^0$. This finishes the definition of primitive $N \times N$-matrix modeled $U(1)$-gauge theory.

Now to define a primitive $N \times N$-matrix modeled $G$-gauge theory one needs to promote from $S(M)_N \twoheadrightarrow M$ to $S(M)_N^V := S(M)_N \otimes V \twoheadrightarrow M$ with $V$ an irreducible representing space of $\mathfrak{g}$, the Lie algebra of $G$. Moreover, the gauge fields as elements of $\mathfrak{g}\otimes\Omega^1_{\text{deR}}(M)$ ($\mathfrak{g}$-valued 1-forms) should be replaced by the elements of $U(\mathfrak{g})\otimes\Omega^1$ ($U(\mathfrak{g})\otimes H$-valued 1-forms) for $U(\mathfrak{g})$ the enveloping algebra of $\mathfrak{g}$. It is well-known that $U(\mathfrak{g})$ is a quantum group and this replacement implies that the matrix modeling approach extends the classical symmetries given by Lie groups to quantum symmetries interpreted by quantum groups. All the structures have been discussed above can be extended to a general primitive $N\times N$-MMG theory by replacing $H$ by $U(\mathfrak{g})\otimes H$ (resp. $\Omega$ with $U(\mathfrak{g})\otimes\Omega$). In fact, $U(\mathfrak{g})\otimes H$ is a $*$-algebra equipped with the induced right $H$-comodule corepresentation $id_{U(\mathfrak{g})}\otimes\varphi$, compatible with its algebraic structure. This makes $U(\mathfrak{g})\otimes H$ to be also an Yetter-Drinfeld module. This comodule structure can naturally be extended to $U(\mathfrak{g})\otimes\Omega$. Moreover, to have $U(\mathfrak{g})\otimes\Omega$ as a DGA one should define the derivative operator d in a compatible sense. It is naturally assumed that d acts only on the components of $\Omega$. It can be seen that it is well defined and converts $U(\mathfrak{g})\otimes\Omega$ to a DGA over $U(\mathfrak{g})\otimes H$. The action of $U(\mathfrak{g})\otimes\Omega$ on $\Gamma(S(M)_N^V)$ is defined according to given triple types (7)-(9) with the difference that in all these module structures the elements of $U(\mathfrak{g})$ act on $\Gamma(S(M)_N^V)$ with;

$$\sum T_{i_1}\otimes\ldots\otimes T_{i_k} \triangleright \psi := \sum T_{i_1}\ldots T_{i_k}\psi,$$

(17)

for any $\sum T_{i_1}\otimes\ldots\otimes T_{i_k} \in U(\mathfrak{g})$ and irrespectively with the action types (7)-(9).

Similarly, $\nabla := d + iA$, $A \in U(\mathfrak{g})\otimes\Omega^1$ defines a connection over $\Gamma(S(M)_N^V)$ with curvature $\hat{R} := idA - A^2$. The Lagrangian density of matter and the Maxwell Lagrangian density (in 4 dimensions) are defined by (11) and (12) respectively. Both $\mathcal{L}_{Matter}$ and $\mathcal{L}_{Maxwell}$ are $U(\mathfrak{g})\otimes\Omega^D$ valued and thus the action (13) is well-defined for non-Abelian cases by extending the trace in (10) to the elements of $U(\mathfrak{g})\otimes H$ naturally.

The similar arguments assert that to have a real valued action, the Hermitian elements of $U(\mathfrak{g})\otimes\Omega^1$ should be considered as the gauge fields. The infinitesimal gauge transformations are defined according to (14) and (15). The action commutator reveals that for a non-Abelian Lie group $G$, with Lie algebra 𝔤, type I and type II acting elements do not commute generally. In fact, similar to the ordinary commutator in the case of standard Yang-Mills theories, the action commutator is the criterion for a MMG theory to be Abelian or non-Abelian. Finally, strait forward calculations assert that to have a gauge invariant action, the infinitesimal gauge transformations ought to be anti-Hermitian elements of $U(\mathfrak{g}) \otimes \Omega^0$. This finishes the definition of primitive MMG theories.

It can be seen that the first approximation of primitive MMG theories lead to commutative ordinary gauge theories with a large number of degrees of freedom which is not of interest. To remove these failures, a constructive conjecture is proposed in the next subsection to decrease the degrees of freedom and to produce the noncommutative gauge theories at the first approximation.

## 2.2. Matrix Modeled Gauge Theories

To cook up a MMG theory in a general framework, one may need some more elaborate structures. The most important case of such structures is the *Matricial Quantization Map* (MQM). An $N \times N$-MQM, $\mathfrak{m}_N$, is in general a ℂ-linear map from $C^\infty(M)$ to $C^\infty(M) \otimes \mathrm{M}_N(\mathbb{C})$ with the following properties;

1. There exists a formal deformation of algebra $C^\infty(M)$ [63-65], shown by $\star$, such that for any set of smooth functions $\{f_i\}_{i=1}^n \in C^\infty(M)$, $n \geq 1$, one finds that;

$$\xi_N tr\{\mathfrak{m}_N(f_1) \ldots \mathfrak{m}_N(f_n)\} = \int_{\mathbb{T}^{d_2}} f_1 \star \ldots \star f_n ,$$

(18)

for some model dependent factor $\xi_N$.

2. $\mathfrak{m}_N$ preserves the exact forms, i.e. for any set of smooth functions $\{f_i\}_{i=1}^n$, $n \geq 1$, and for any $1 \leq \mu \leq D$, there exists a smooth function $f$ such that;

$$\sum_{i=1}^{n} tr\{\mathfrak{m}_N(f_1) \ldots \mathfrak{m}_N(\partial_\mu f_i) \ldots \mathfrak{m}_N(f_n)\} = \int_{\mathbb{T}^{d_2}} \partial_\mu f .$$

(19)

3. For any two sets of smooth functions $\{f_j^i\}_{i,j=1}^{n,n_i}$ and $\{g_j^i\}_{i,j=1}^{m,m_i}$, the equality $\sum_{i=1}^n \mathfrak{m}_N(f_1^i)\mathfrak{m}_N(f_2^i) \ldots \mathfrak{m}_N(f_{n_i}^i) = \sum_{i=1}^m \mathfrak{m}_N(g_1^i)\mathfrak{m}_N(g_2^i) \ldots \mathfrak{m}_N(g_{m_i}^i)$ yields

$$\sum_{i,k=1}^{n,n_i} \mathfrak{m}_N(f_1^i) \ldots \mathfrak{m}_N(\partial_\mu f_k^i) \ldots \mathfrak{m}_N(f_{n_i}^i) = \sum_{i,k=1}^{m,m_i} \mathfrak{m}_N(g_1^i) \ldots \mathfrak{m}_N(\partial_\mu g_k^i) \ldots \mathfrak{m}_N(g_{m_i}^i)$$

(20)

for any $1 \leq \mu \leq D$.

One may usually combine the first and the second properties of MQM $\mathfrak{m}_N$ to give an alternative formulation for the second property;

2*. For any set of smooth functions $\{f_i\}_{i=1}^n$, $n \geq 1$, and for any $1 \leq \mu \leq D$, one finds;

$$\sum_{i=1}^{n} \int_{\mathbb{T}^{d_2}} f_1 \star \ldots \star \partial_\mu f_i \ldots \star f_n = \int_{\mathbb{T}^{d_2}} \partial_\mu f ,$$

(21)

for some $f \in C^\infty(M)$.

It may also be assumed that a MQM sends any real $f \in C^\infty(M)$ to a Hermitian matrix valued function. It is also supposed that the entries of $\mathfrak{m}_N$ are pseudo-differential operators and thus the entries of $\mathfrak{m}_N(f)(x)$ dependant only on the germ of $f$ at $x$. On the other hand, $\mathfrak{m}_N$ may or may not keep the algebraic structure, but it can be shown that algebra morphism MQMs are not of interest. Indeed, it is seen in the following that, if $\mathfrak{m}_N$ be an algebra homomorphism, then the matrix modeling would lead to ordinary Yang-Mills theories.

Assume that an $N \times N$-MQM, say $\mathfrak{m}_N$, is given. For any smooth function $f$, one may conventionally denote $\mathfrak{m}_N(f)$ by $[f]_N$. Let $\mathcal{H}$ be the tensor algebra generated by $\text{Im}(\mathfrak{m}_N)$. Indeed $\mathcal{H}$ is a Hopf sub-algebra of $H$ with induced structures. More precisely, since the inclusion map, $i: \mathcal{H} \hookrightarrow H = \Omega^0$, defines an algebra homomorphism, then $(\mathcal{H}, \Omega, \int)$ would be a $D$-dimensional cycle over $\mathcal{H}$ [7]. But unfortunately this cycle has no more interesting properties than $(H, \Omega, \int)$. Indeed they both lead to the same gauge theories in general. In fact $\Omega^d = H \otimes \Omega^d_{\text{deR}/\mathcal{SP}}(M)$, $d \geq 1$, while one may be interested in $\mathcal{H} \otimes \Omega^d_{\text{deR}/\mathcal{SP}}(M)$. To overcome this failure it should be defined a new fashioned DGA over $\mathcal{H}$. This can be done by modifying the action of $\mathcal{T}(M)$ on $\mathcal{H}$. It can be easily seen that $(\mathcal{H}, \mathcal{T}(M), \tau_\mathfrak{m})$, with $\tau_{\mathfrak{m}_{X_t}}([f]_N) := [f \circ X_t]_N$, $X \in \text{Lie}\, \mathcal{T}(M)$, $t \in \mathbb{R}$, is a dynamical system. This naturally induces a representation of $\text{Lie}\, \mathcal{T}(M)$ in the Lie algebra of derivations of $\mathcal{H}$, called the *matrix modeling derivation*, $d_\mathfrak{m}$. Thus $d_\mathfrak{m}$ is a $\mathbb{C}$-linear map from $\mathcal{H}$ to $\mathcal{H} \otimes (\text{Lie}\, \mathcal{T}(M))^*$ which vanishes on the unity and acts on the generators of $\mathcal{H}$ by $d_\mathfrak{m}[f]_N = [\partial_\mu f]_N dx^\mu$, $f \in C^\infty(M)$. To see this in details, extend $d_\mathfrak{m}$ to all of $\mathcal{H}$ as a derivation to build a first order differential calculus on $\mathcal{H}$ [10, 11, 12, 20]. Set $\Xi^d = \mathcal{H} \otimes \Omega^d_{\text{deR}/\mathcal{SP}}(M)$, $d \geq 0$. Clearly $d_\mathfrak{m}$ can be naturally extended to $\Xi^d$ for any $1 \leq d \leq D$. It can be checked that $d_\mathfrak{m}$ is well-defined over $\mathcal{SP}$-class coordinate systems. It is seen easily that $d_\mathfrak{m}$ is nilpotent, i.e. $d_\mathfrak{m}^2 = 0$. Considering $\Xi = \oplus_d \Xi^d$ as a graded algebra, it can also be shown that $d_\mathfrak{m}$ obeys the graded Leibnitz rule, i.e.;

$$d_\mathfrak{m}(\omega_1 \omega_2) = d_\mathfrak{m}\omega_1 \omega_2 + (-)^{|\omega_1|} \omega_1 d_\mathfrak{m}\omega_2,$$

(22)

for any two elements $\omega_1, \omega_2 \in \Xi$. Thus $(\Xi, d_\mathfrak{m})$ defines a differential calculus over $\mathcal{H}$. Moreover, it can be seen that the integration of $(H, \Omega, \int)$ naturally induces an integration structure over $(\Xi, d_\mathfrak{m})$. Indeed;

$$\int \sum a_{i_1} \otimes \ldots \otimes a_{i_k} \, dx^1 \wedge \ldots \wedge dx^D := \sum \int_{\mathbb{R}^{d_1}} \xi_N tr\{a_{i_1} \ldots a_{i_k}\},$$

(23)

with $\sum a_{i_1} \otimes \ldots \otimes a_{i_k} \in \mathcal{H}$, defines a $D$-dimensional integration structure over $\Xi$. It is easily seen that; $\int \omega_1 \omega_2 = (-)^{|\omega_1||\omega_2|} \int \omega_2 \omega_1$, for $\omega_1, \omega_2 \in \Xi$. But (21) implies that $\int d_m \omega = 0$ for all $\omega \in \Xi$. Therefore, $(\mathcal{H}, \Xi, \int)$ is a $D$-dimensional cycle over $\mathcal{H}$. $\Xi$ also acts on $\Gamma(S(M)_N)$ in all three types (7)-(9).

Now one needs a connection structure over $\Gamma(S(M)_N)$ as a $\mathcal{H}$-module. Although $\Gamma(S(M)_N)$ is not a finitely generated projective module, but the third property of MQMs, lets one to construct a well-defined covariant derivation on $\Gamma(S(M)_N)$. It is enough to assume that $\Gamma(S(M)_N)$ is spammed with elements; $1 \otimes e_k$ and $\mathfrak{m}_N(f_1) \ldots \mathfrak{m}_N(f_n) \otimes e_k$, $n \in \mathbb{N}$, over $\mathbb{C}$ for $\{e_k\}_{k=1}^{2^{[D/2]}}$ a global parallel basis for $S(M)$. Therefore, by this assumption and according to the third property of MQMs, the matrix modeling derivation $d_m$, naturally induces a flat connection over $\Gamma(S(M)_N)$ also denoted by $d_m$. Moreover, following [78], any $A \in \Xi^1$ defines a connection over $\Gamma(S(M)_N)$ by $\nabla = d_m + iA$ with curvature $\hat{R} = id_m A - A^2 \in \Xi^2$.

Now it is the time to define the Lagrangian density for an $N \times N$-matrix modeled $U(1)$-gauge theory. Since $\mathcal{H} \subseteq H$, what were constructed for primitive MMG theories can be used for MMG theories. Indeed, the Lagrangian of matter and the Maxwell Lagrangian (in 4 dimensions) are similarly defined by (11) and (12) respectively while the action is given by integration (23). Analogously the Hermitian elements of $\Xi^1$ can be considered as the gauge fields while the anti-Hermitian elements of $\Xi^0$ are used as the infinitesimal gauge transformations. Note that $\mathcal{H}$ doesn't admits the involution structure of $H$, generally. Actually, for any element $a \in \mathcal{H}$, $a^*$ is not necessarily an element of $\mathcal{H}$. But the given formalism never suffers by this. In fact, $a \in \mathcal{H}$ is defined to be a (anti-) Hermitian element if and only if it is (anti-) Hermitian as an element of $H$.

As it was discussed for primitive MMG theories, to define an $N \times N$-matrix modeled $G$-gauge theory, for a general non-Abelian semi-simple Lie group $G$ with algebra $\mathfrak{g}$, one should respectively replace $\mathcal{H}$ (resp. $\Xi$) and $S(M)_N$ by $U(\mathfrak{g}) \otimes \mathcal{H}$ (resp. $U(\mathfrak{g}) \otimes \Xi$) and $S(M)_N^V$, for $V$ an irreducible representing space of $G$. The details of the formalism go on as mentioned for primitive MMG theories.

Working out the action of an arbitrary MMG theory leads to primary intuitions about matrix modeling. At the first step the Abelian theories are discussed. To write down the action for an Abelian MMG theory one should consider a number of assumptions for simplicity. The first assumption is about the gauge fields, say for instance; $A = A_\mu dx^\mu \in \mathcal{H} \otimes \Omega^1_{\text{deR}/\mathcal{SP}}(M)$. It is supposed that $A_\mu$s for all $\mu$ are homogeneous elements of degree 1, i.e. for all $\mu$, $A_\mu \in \text{Im } \mathfrak{m}_N = \mathfrak{m}_N(C^\infty(M, \mathbb{R}))$. Thus, conventionally the gauge fields, $A_\mu$, are shown by $[A_\mu]_N$. By this, the curvature is given by;

$$\hat{R} = \frac{1}{2}(i[\partial_\mu A_\nu - \partial_\nu A_\mu]_N - [[A_\mu]_N, [A_\nu]_N]_\otimes)dx^\mu \wedge dx^\nu,$$

(24)

with $[,]_\otimes$ is the commutator for tensor product, $\otimes$. Actually $[,]_\otimes$ is a special case of $[,]_{\text{Act.}}$, but here it is used to insist on the action independent soul of the curvature $\hat{R}$. Therefore, for all action types (7)-(9), the Maxwell action of an Abelian $N \times N$-MMG theory is given by;

$$S_{Maxwell} = \int_{\mathbb{R}^{d_1} \times \mathbb{T}^{d_2}} -\frac{1}{4} F_{\mu\nu} \star F^{\mu\nu},$$

(25)

where $F_{\mu\nu} = \partial_\mu A_\nu - \partial_\nu A_\mu + i[A_\mu, A_\nu]_\star$ for $[,]_\star$ the commutator for star product $\star$. Note that according to the first assumption, action (25) can be considered as the first approximation of pure Abelian MMG theory.

According to (25) one can consider that;

$$\mathcal{L}_{Maxwell} = -\frac{1}{4} F_{\mu\nu} \star F^{\mu\nu},$$

(26)

and consequently find a noncommutative pure gauge theory equivalent to the pure Abelian MMG theory at the first approximation.

The second assumption is about matricial spinor fields, say $\psi \in \Gamma(S(M)_N)$. According to the assumed structure of $\Gamma(S(M)_N)$ one may use the simplest form for $\psi$ by considering $\psi \in \mathfrak{m}_N(C^\infty(M,\mathbb{R})) \otimes \mathbb{C}^{2^{[D/2]}}$, i.e. $\psi = [\psi_k]_N \otimes e_k$, $k = 1, \ldots, 2^{[D/2]}$, for a number of smooth functions $\psi_k$ and for $\{e_k\}_{k=1}^{2^{[D/2]}}$ a global parallel basis for $S(M)$. Thus similarly the spinor $\psi$, is shown by $[\psi]_N$. Despite of the Maxwell action, the module structure of $\Gamma(S(M)_N)$ described by the actions (7)-(9), plays a crucial role in the matter action. Indeed, each of the action types I, II and III, will yield a particularly distinct theory.

As mentioned above, the action of matter is defined in term of the integration structure of cycle $(H, \Omega, \int\,)$. Therefore, splitting $S_{Matter}$ into the free term, $S^0_{Matter}$, defined by the flat connection $d_m$, and the interaction part, $S^{int}_{Matter}$, one finds that;

$$S^0_{Matter} = \int_{\mathbb{R}^{d_1} \times \mathbb{T}^{d_2}} i\gamma^\mu_{ab} \bar{\psi}_a \star \partial_\mu \psi_b \,,$$

(27)

for all three types of module structures (7)-(9).

On the other hand, the interaction term of the matter action, for each type of module structure (7)-(9), is respectively given by;

Type I )

$$S^{int}_{I} = -\int_{\mathbb{R}^{d_1} \times \mathbb{T}^{d_2}} \gamma^\mu_{ab} \bar{\psi}_a \star A_\mu \star \psi_b \,.$$

(28)

Type II )

$$S^{int}_{II} = \int_{\mathbb{R}^{d_1} \times \mathbb{T}^{d_2}} \gamma^\mu_{ab} \bar{\psi}_a \star \psi_b \star A_\mu \,.$$

(29)

Type III )

$$S_{III}^{int} = -\int_{\mathbb{R}^{d_1} \times \mathbb{T}^{d_2}} \gamma_{ab}^{\mu} \bar{\psi}_a \star [A_\mu, \psi_b]_\star = S_I^{int} + S_{II}^{int}.$$

(30)

Note that according to the first and the second assumptions, (27)-(30) give hand the firs approximation of matter action of an Abelian MMG theory.

Similar to the case of pure gauge theories one can equivalently define;

$$\mathcal{L}_{Matter} = i\gamma_{ab}^{\mu} \bar{\psi}_a \star \partial_\mu \psi_b - \gamma_{ab}^{\mu} \bar{\psi}_a \star A_\mu \star \psi_b,$$

(31)

to gain a noncommutative gauge theory equivalent to the type I Abelian MMG theory at the first approximation. Indeed, (25)-(31) define an equivalence relation between Abelian noncommutative gauge theories and the first approximation of Abelian MMG theories. In the other words, Abelian MMG theories produce the noncommutative Abelian gauge theories in the first approximation. Such an equivalence relation can also be found for non-Abelian cases. To derive the simplest form of the action of non-Abelian MMG theories one needs the third assumption. In fact, for simplicity one should assume that the gauge fields $A \in U(\mathfrak{g}) \otimes \Xi^1$, are given in the forms of $T^a \otimes [A_\mu^a]_N dx^\mu$, for $\{T^a\}_{a=1}^{\dim \mathfrak{g}}$ a Hermitian basis of $\mathfrak{g}$. According to these assumptions the first approximation of a pure non-Abelian MMG theory is given by the following Maxwell's action;

$$S_{Maxwell} = \int_{\mathbb{R}^{d_1} \times \mathbb{T}^{d_2}} -\frac{1}{4} tr\{F_{\mu\nu} \star F^{\mu\nu}\},$$

(32)

for $F_{\mu\nu} = (\partial_\mu A_\nu^a - \partial_\nu A_\mu^a - C^{abc} A_\mu^b \star A_\nu^c) T^a + i[A_\mu^a, A_\nu^b]_\star T^b T^a$, with $iC^{abc}$ the structure constants of the given Hermitian basis $\{T^a\}_{a=1}^{\dim \mathfrak{g}}$. The free part of the matter action is equal to (27), but the interaction terms, with respect to the module structures, are given by;

Type I )

$$S_{\text{I}}^{int} = -\int_{\mathbb{R}^{d_1} \times \mathbb{T}^{d_2}} T_{ij}^a \, \gamma_{\alpha\beta}^{\mu} \, \bar{\psi}_{\alpha,i} \star A_{\mu}^a \star \psi_{\beta,j} \, .$$

(33)

Type II )

$$S_{\text{II}}^{int} = \int_{\mathbb{R}^{d_1} \times \mathbb{T}^{d_2}} T_{ij}^a \, \gamma_{\alpha\beta}^{\mu} \, \bar{\psi}_{\alpha,i} \star \psi_{\beta,j} \star A_{\mu}^a \, .$$

(34)

Type III )

$$S_{\text{III}}^{int} = -\int_{\mathbb{R}^{d_1} \times \mathbb{T}^{d_2}} T_{ij}^a \, \gamma_{\alpha\beta}^{\mu} \, \bar{\psi}_{\alpha,i} \star [A_{\mu}^a, \psi_{\beta,j}]_{\star} = S_{\text{I}}^{int} + S_{\text{II}}^{int} \, .$$

(35)

Similarly one can set;

$$\mathcal{L}_{Maxwell} = -\frac{1}{4} tr\{F_{\mu\nu} \star F^{\mu\nu}\},$$

(36)

and;

$$\mathcal{L}_{Matter} = i\gamma_{ab}^{\mu} \bar{\psi}_a \star \partial_{\mu} \psi_b + T_{ij}^a \, \gamma_{\alpha\beta}^{\mu} \, \bar{\psi}_{\alpha,i} \star \psi_{\beta,j} \star A_{\mu}^a \, ,$$

(37)

to define an equivalent noncommutative non-Abelian gauge theory for the first approximation of non-Abelian MMG theory of type II. Indeed, this shows that MMG theories are equivalent to noncommutative gauge theories at the first approximation.

Although it may seem that no new achievements were worked out during this section but in the following it is shown that the matrix modeling formulation gives a commutative spirit to diagram calculations in noncommutative gauge theories which leads to a great simplification of manipulations. Moreover, the matrix modeling formulation produces an elaborate framework to interpret the noncommutative gauge theories in the setting of noncommutative geometry. This elegant property helps one to extract a number of nonperturbative results for noncommutative gauge theories.

Before closing this section a number of remarks should be given. The first remark is that the matrix modeling procedure has been given above, can be similarly used to interpret noncommutative scalar field theories in matricial formulation which is of interest by itself. On the other hand, one can go farther and define similarly $N_1 \times N_2$-MMG theories, $N_1 \neq N_2$, which may need its own technicalities. Second, the axial extension of gauge theories can be similarly worked out for even dimensional MMG theories [80]. To this end it is enough to replace $\mathcal{H}$ with $\mathcal{H} \oplus \gamma_5 \mathcal{H}$ as an algebra and follow the given procedure analogously. The details are similar to [80]. Next, it can be seen that to define non-Abelian MMG theories, $U(\mathfrak{g})$ can be replaced by any arbitrary quantum group. This may results in new definition of gauge theories based on quantum groups with or without underlying group structures. The most convenient cases are quantum groups $U_q(\mathfrak{sl}_2)$ and $SL_q(2)$. Also $U_q(\mathfrak{sl}_2)$ can be replaced by $U_q(\mathfrak{g})$ for any semi-simple Lie algebra $\mathfrak{g}$ as a 1-parameter deformation of $U(\mathfrak{g})$. Finally, it should be cleared that in the spirit of defining the MQMs, one is solely looking for a deformation quantization structure or equivalently a star product, $\star$, over $C^\infty(M)$ which is *tracial* (the property 1) and *exact* (the property $2^*$). As will be discussed in the following translation-invariant products are exact by definition. On the other hand, any translation-invariant product is substantially tracial [61]. Therefore in the following the translation-invariant noncommutative gauge (TNG) theories are of extreme interest.

By Kontsevich's quantization formula [81], any Poisson bracket over $C^\infty(M)$ with property $\partial_\mu \{f, g\} = \{\partial_\mu f, g\} + \{f, \partial_\mu g\}$, $f, g \in C^\infty(M)$, for a densely defined coordinate system $\{x^\nu\}_{\nu=1}^D$ and for any $\mu = 1, \dots, D$, leads to an exact deformation quantizetion of $C^\infty(M)$. Thus, to defining a translation-invariant star product one may look for symplectic structures on $M$ which admit densely defined Darboux's charts [82]. The most natural quantization struct-ure occurs when $d_1 = d_2 = d$. In this case $M = T^*\mathbb{T}^d$ and thus it admits a natural symplectic structure. The canonical Poisson bracket due to this symplectic form leads to a Groenewold-Moyal star product over the set of Schwartz functions on $M$ which is obviously translation-invariant. Although it seems to be the most natural quantization structure on $M$, but here it is more desirable to deal with those Groenewold-Moyal star products which leave noncompact coordinates commutative. Also one may be interested in general forms of translation-invariant star products which may admit no symplectic structures.

## 3. Matrix Modeling and Translation-Invariant Noncommutative Gauge Theories

As mentioned in the introduction, the main goal of constructing the elaborate formalism of MMG theories is their relationship to noncommutative Yang-Mills (NCYM) theories. In this section, it is shown that the conjecture of MQMs comes true as the limit of a series of matricial formulations for any translation-invariant star product. Therefore, roughly speaking, any TNG theory can be considered as an asymptotic theory of an appropriate family of MMG theories. Indeed it is shown that for any translation-invariant noncommutative $U(\mathfrak{n})$-gauge theory, there exists a collection of matrix formulations $\mathfrak{m}_N$, $N \in \mathbb{N}$, approximately satisfying the properties of MQMs, and leading to a family of compatible $N \times N$-MMG theories which explain the theory asymptotically for large $N$ limits. In the other words, MMG theories can be considered as the effective theories of TNG theories. On the other hand, it is shown that in the sense of Groenewold-Moyal noncommutative gauge theories, the matrix modeling approach leads to a $\theta$-expanded formulation of noncommutative gauge theories in matricial settings. Actually, it is seen that MQMs give rise to a generalization of Seiberg-Witten map [40] for translation-invariant noncommutative fields in matrix model settings.

## 3.1. Translation-Invariant Products

A noncommutative structure on $\mathbb{R}^{2n}$, is usually given by nontrivial commutation relations of coordinate functions of a fixed globally defined chart, say $(\mathbb{R}^{2n}, \{x^\mu\}_{\mu=1}^{2n})$, with an anti-symmetric constant matrix $\theta$;

$$[x^\mu, x^\nu] = i\theta^{\mu\nu}, \tag{38}$$

$\theta^{\mu\nu} = -\theta^{\nu\mu} \in \mathbb{C}$. It is seen that commutation relations (38) can be satisfied by replacing the ordinary product of $C^\infty(\mathbb{R}^{2n})$ by a noncommutative one, $\star$;

$$x^\mu \star x^\nu - x^\nu \star x^\mu = i\theta^{\mu\nu}. \tag{39}$$

Usually the quantization structure $\star$ is considered as the well-known Groenewold-Moyal product with $f \star g \coloneqq \pi(e^{\theta^{\mu\nu}\partial_\mu \otimes \partial_\nu}(f \otimes g))$ for $\theta$ the constant Hermitian matrix proportional to $i\begin{pmatrix} 0 & 1 \\ -1 & 0 \end{pmatrix}$ for any 2-dimensional noncommutative subspace and for $\pi: C^\infty(\mathbb{R}^{2n}) \otimes C^\infty(\mathbb{R}^{2n}) \to C^\infty(\mathbb{R}^{2n})$ the ordinary point-wise production. The simplest generalization of Groenewold-Moyal product is the Wick-Voros production [83-86] which is defined similar to the Groenewold-Moyal one with replacing $i\begin{pmatrix} 0 & 1 \\ -1 & 0 \end{pmatrix}$ by $\begin{pmatrix} 1 & i \\ -i & 1 \end{pmatrix}$. It can be seen that Groenewold-Moyal and Wick-Voros products are generally well defined over $\mathcal{S}(\mathbb{R}^{2n})$, the Schwartz class of functions. It can also be shown that these two star products both can be regarded as deformation quantization due to Weyl-Wigner correspondence [87-89].

More than constant commutation relation (38), there may be defined other deformation structures on $\mathbb{R}^{2n}$ with linear and quadratic forms. The linear case leads to a Lie algebra with;

$$[x^\mu, x^\nu] = i\lambda^{\mu\nu}{}_\sigma x^\sigma, \tag{40}$$

$\lambda^{\mu\nu}{}_\sigma \in \mathbb{C}$. These structures are basically discussed in two different settings, fuzzy spaces [90, 91] and $\kappa$-deformation [92, 93]. The quadratic commutation relations are given in terms of $R$-matrix or quasi-triangular structures of quantum groups [94];

$$[x^\mu, x^\nu] = \left(\frac{1}{q} R^{\mu\nu}_{\sigma\lambda} - \delta^\mu_\lambda \delta^\nu_\sigma\right) x^\sigma x^\lambda .$$

(41)

Indeed, $R$-matrices are the solutions of quantum Yang-Baxter equation in quantum inverse scattering theory [17-20]. $R$-matrix structure enables one to formulate the quantum statistics in the setting of fusion theory in modular tensor categories [95]. Actually, $R$-matrix structure permits a quantum group to give representations of braid groups as a foundation of quantum statistical mechanics. The representations are given in terms of Reidemeister moves for links and knots [18, 19, 76]. Finally, the ribbon structure as a special kind of quasi-triangular structures, enables a quantum group to produce a modular category over its irreducible representing spaces as simple objects [19, 75]. This produces the base of mathematics for studying $2+1$-dimensional Chern-Simons gauge theories for anyons [95].

It is seen that noncommutative structure (39), despite of (40) and (41), is independent of coordinate functions $x^\mu$s. Such deformation structures are called *translation-invariant products*. Translation-invariant products enable noncommutative gauge theories to be translation-invariant and consequently to preserve the energy-momentum conservation law. Therefore, translation-invariant noncommutative products play a crucial role in the realm of researches of noncommutative field theories [61, 62].

More precisely a star product on $C^\infty(\mathbb{R}^m)$, is translation-invariant if;

$$\mathcal{T}_a(f) \star \mathcal{T}_a(g) = \mathcal{T}_a(f \star g),$$

(42)

for any vector $a \in \mathbb{R}^m$ and for any $f, g \in C^\infty(\mathbb{R}^m)$, where $\mathcal{T}_a$, is the translating operator; $\mathcal{T}_a(f)(x) = f(x + a)$, $f \in C^\infty(\mathbb{R}^m)$. Replacing $a$ with $ta$, $t \in \mathbb{R}$, in

(42) and differentiating with respect to $t$ at $t = 0$, one easily finds that $\partial_\mu (f \star g) = \partial_\mu f \star g + f \star \partial_\mu g$, which shows that any translation-invariant product is exact. To be precautious and to have well-defined products, from now on $C^\infty(\mathbb{R}^m)$ is replaced by $\mathcal{S}(\mathbb{R}^m)$ for any translation-invariant product $\star$.

An equivalent definition of translation-invariant products over Cartesian space $\mathbb{R}^m$, is given by [61];

$$(f \star g)(x) := \int \frac{\mathrm{d}^m p}{(2\pi)^m} \frac{\mathrm{d}^m q}{(2\pi)^m} \tilde{f}(q) \tilde{g}(p) e^{\alpha(p+q,p)} e^{i(p+q).x} ,$$

(43)

for $f, g \in \mathcal{S}(\mathbb{R}^m)$, their Fourier transformations $\tilde{f}, \tilde{g} \in \mathcal{S}(\mathbb{R}^m)$, and finally for a 2-cycle $\alpha \in C^\infty(\mathbb{R}^m \times \mathbb{R}^m)$ which obeys the following cyclic property;

$$\alpha(p, r+s) + \alpha(r+s, r) = \alpha(p, r) + \alpha(p-r, s) ,$$

(44)

for any $p, r, s \in \mathbb{R}^m$. Actually (44) is equivalent to associativity of $\star$. In fact, (44) is reflected by the condition $(f \star g) \star h = f \star (g \star h)$ for any three Schwartz functions $f, g$ and $h$. On the other hand, it can be easily checked that the definition (43) defines a translation-invariant product in agreement with definition (42). Thus, (43) is the most general definition for translation-invariant deformation quantization of $C^\infty(\mathbb{R}^m)$. For translation-invariant deformation structures of $C^\infty(\mathbb{T}^m)$, the integrations in (43) should be replaced by discrete summations on the lattice of Fourier modes. It can also be checked that for any $f \in \mathcal{S}(\mathbb{R}^m)$, $1 \star f = f \star 1 = f$ if and only if;

$$\alpha(p, p) = \alpha(p, 0) = 0 ,$$

(45)

for any $p \in \mathbb{R}^m$. Moreover, combining (41) and (42) leads to;

$$\alpha(0, p) = \alpha(0, -p) ,$$

(46)

for any $p \in \mathbb{R}^m$. Using (46) it can be shown that any translation-invariant product admits the trace property;

$$\int_{\mathbb{R}^m} f_1 \star \ldots \star f_{k-1} \star f_k = \int_{\mathbb{R}^m} f_k \star f_1 \star \ldots \star f_{k-1}$$

(47)

for any $k \in \mathbb{N}$ and for any set of $f_1, f_2, \ldots, f_k \in \mathcal{S}(\mathbb{R}^m)$. Indeed;

$$\int_{\mathbb{R}^m} f \star g = \int d^m x \frac{d^m p}{(2\pi)^m} \frac{d^m q}{(2\pi)^m} \tilde{f}(q) \tilde{g}(p) e^{\alpha(p+q,p)} e^{i(p+q).x}$$

$$= \int \frac{d^m p}{(2\pi)^m} \tilde{f}(-p) \tilde{g}(p) e^{\alpha(0,p)} = \int \frac{d^m p}{(2\pi)^m} \tilde{g}(-p) \tilde{f}(p) e^{\alpha(0,p)}$$

$$= \int d^m x \frac{d^m p}{(2\pi)^m} \frac{d^m q}{(2\pi)^m} \tilde{g}(q) \tilde{f}(p) e^{\alpha(p+q,p)} e^{i(p+q).x}$$

$$= \int_{\mathbb{R}^m} g \star f$$

(48)

for any $f, g \in \mathcal{S}(\mathbb{R}^m)$. Therefore, (47) follows by induction. Also from (44)-(46) it can be shown that;

$$\alpha(p, q) = -\alpha(q, p) + \alpha(0, q - p),$$

(49)

$$\alpha(0, q) = \alpha(0, p) - \alpha(q, p) + \alpha(-p, q - p),$$

(50)

and finally

$$\alpha(p, q) = -\alpha(0, p) + \alpha(0, q) + \alpha(0, p - q) - \alpha(-p, q - p),$$

(51)

for any $p, q \in \mathbb{R}^m$. It can also be checked that the commutivity of $\star$ is equivalent to;

$$\alpha(p, q) = \alpha(p, p - q),$$

(52)

for any $p, q \in \mathbb{R}^m$. Therefore, $\alpha$ is called commutative if it satisfies (52).

### 3.2. Matrix Modeling and Translation-Invariant Products

Now we are ready to study TNG theories with more details. One of the standard definitions of TNG theories over $\mathbb{R}^{2n}$ is given by the following Lagrangian density [40, 47];

$$\mathcal{L}(\psi) = \bar{\psi}_{\alpha,i} \star i\partial_\mu \psi_{\beta,i} \gamma^\mu_{\alpha\beta} - T^a_{ij} \gamma^\mu_{\alpha\beta} \bar{\psi}_{\alpha,i} \star A^a_\mu \star \psi_{\beta,j},$$

(53)

with $\star$ a translation-invariant product over $C^\infty(\mathbb{R}^{2n})$ and $T^a$s the Hermitian matrices as the represented generators of the Lie algebra. Obviously, for noncommutative QED the color matrices $T^a$, disappears in (53). On the other hand, the noncommutativity of $\star$ gives rise to a number of different definitions for TNG theories. For instance a natural alternative for definition (53) is;

$$\mathcal{L}(\psi) = \bar{\psi}_{\alpha,i} \star i\partial_\mu \psi_{\beta,i} \gamma^\mu_{\alpha\beta} + T^a_{ij} \gamma^\mu_{\alpha\beta} \bar{\psi}_{\alpha,i} \star \psi_{\beta,j} \star A^a_\mu.$$

(54)

The expression (54) of NCYM theories, originally contains the currents, while the currents only appear in the action of (54) due to the trace property of $\star$. On the other hand, NCYM theories can be also defined in another form;

$$\mathcal{L}_{int}(\psi) = -T^a_{ij} \gamma^\mu_{\alpha\beta} \bar{\psi}_{\alpha,i} \star [A^a_\mu, \psi_{\beta,j}]_\star.$$

(55)

Equations (33)-(35) show that all the Lagrangian densities of (53)-(55) can be unified in the formalism of matrix modeling if a MQM is found for $\star$. It is claimed that for any translation-invariant product $\star$, there exist a family of matrix formulations, say $\{\mathfrak{m}_N\}_{N\in\mathbb{N}}$, approximately satisfying the properties of MQMs, leading to a family of star products, $\{\star_N\}_{N\in\mathbb{N}}$, which $\star_N$ tends to $\star$ uniformly, as $N \to \infty$. To show this fact, a list of notations is set initially;

- a) Let $(\mu) = (\mu_1, \dots, \mu_m)$ be an $m$-plet of integers. Then, set $|\mu| = m$, the length of $(\mu)$. For the empty multi-plet $(\emptyset) = (\ )$, set $|\emptyset| = 0$.
- b) Assume that $k \in \mathbb{R}^{2n}$ is an arbitrary vector and $(\mu)$ is a multi-plet with $|\mu| = m$. Then, by convention, $k^{(\mu)}$ means $k^{\mu_1} \times \dots \times k^{\mu_m}$. One also needs $k^{(\emptyset)}$ to be equal to unity for any $k \in \mathbb{R}^{2n}$. Moreover, one can naturally generalize this notation for any arbitrary tensor with finitely many up and down indices. For example if $\Lambda$ is a matrix with up indices and $(\mu)$ and $(\nu)$ are two arbitrary multi-plets, then;

$$\Lambda^{(\mu)(\nu)} = \begin{cases} \Lambda^{\mu_1\nu_1} \times \dots \times \Lambda^{\mu_m\nu_m} & \text{if } |\mu| = |\nu| = m \\ 0 & \text{otherwise} \end{cases}.$$

(56)

Also one demands for $\Lambda^{(\emptyset)(\emptyset)} = 1$ similarly.

- c) As was stated above, the space-time manifold is the Cartesian product space $\mathbb{R}^{d_1} \times \mathbb{T}^{d_2}$. Here it is suppose that $d_2 = 2d$ and $D = d_1 + d_2 = 2n$. Indeed, we consider that $M = \mathbb{R}^{2(n-d)} \times \mathbb{T}^{2d}$. It is also supposed that the radius $R$ of $\mathbb{T}^{2d}$ tends to infinity; $R \to \infty$. As it is seen in the following, for the case of Groenewold-Moyal noncommutative fields, $\mathbb{T}^{2d}$ plays the role of noncommutative torus $\mathbb{T}^{2d}_\theta$ for TNG theories in large $N$ limit of matrix models.
- d) Consider $\alpha$ from (43). Set $\alpha(p+q, p) = \langle \alpha_L(ip) | \alpha_R(iq) \rangle$, $p, q \in \mathbb{R}^{2n}$, for two either finite or possibly infinite dimensional vector valued functions $|\alpha_L\rangle$ and $|\alpha_R\rangle$. Actually the Taylor expansion formula for $\alpha$ with radius of convergence $\infty$, produces such vector valued functions. Thus, for infinite dimensional vectors $|\alpha_L\rangle$ and $|\alpha_R\rangle$, it should be supposed that the entries $\alpha_L^k$ and $\alpha_R^k$, tend to zero rapidly as $k \to \infty$. Here for rapidly decaying we assumed that; $\lim_{k\to\infty} k^s \alpha_L^k(ip) \alpha_R^k(iq) = 0$ for any $p, q \in \mathbb{R}^{2n}$ and for any $s \in \mathbb{N}$. On the other hand, (45) implies that; $|\alpha_L(0)\rangle = |\alpha_R(0)\rangle = 0$.

- e) Eventually for any $r \in \mathbb{N}$ and to any given vector $|\alpha\rangle$, we correspond an $r$-dimensional vector $|\alpha^{(r)}\rangle$ which its entries coincides with the first $r$ entries of $|\alpha\rangle$.

Now suppose that $m \geq 0$ is an integer and set $N = \frac{(m)^{m+1}-1}{m-1}(2m+1)^{2d}$. To define the MQM $\mathfrak{m}_N$ for translation-invariant product (43), one should use the Fourier modes over the torus $\mathbb{T}^{2d}$. For any $f \in \mathcal{S}(\mathbb{T}^{2d} \times \mathbb{R}^{2(n-d)})$ and for any two Fourier modes $\vec{p} = (p_1, \ldots, p_{2d}), \vec{q} = (p_1, \ldots, p_{2d}) \in \mathbb{Z}^{2d}$, set;

$$[f]_{N\vec{p}\vec{q}}^{(\mu)(\nu)}(x) := \frac{\alpha_R^{(m)}(ik_{\vec{p}})^{(\mu)} \alpha_L^{(m)}\left(i(k_{\vec{p}} - k_{\vec{q}})\right)^{(\nu)}}{\sqrt{|\mu|!} \sqrt{|\nu|!}} \hat{f}(x, k_{\vec{p}} - k_{\vec{q}}),$$

(57)

with $k_{\vec{p}} = \frac{2\pi}{2\pi R}\vec{p} = \frac{1}{R}\vec{p}$, $x \in \mathbb{R}^{2(n-d)}$. Where $\hat{f}(x, k_{\vec{p}} - k_{\vec{q}})$ is the Fourier transformation of $f$ over $\mathbb{T}^{2d}$;

$$\hat{f}(x, k_{\vec{p}}) = \int_{y \in \mathbb{T}^{2d}} f(x, y) \frac{e^{-ik_{\vec{p}} \cdot y}}{(2\pi R)^{2d}}.$$

(58)

Note that the Haar measure on $\mathbb{T}^{2d}$ is chosen so that the volume of $\mathbb{T}^{2d}$ equals to unity. To keep the matrix of (57) finite dimensional, one has to restrict the Fourier modes of $\mathbb{T}^{2d}$; $\max(\{|p_i|\}_{i=1}^{2d} \cup \{|q_i|\}_{i=1}^{2d}) \leq m$. Also it is supposed that $|\mu|, |\nu| \leq m$. Therefore, $[f]_N$ is an $N \times N$ matrix valued Schwartz function over $\mathbb{R}^{2(n-d)}$. Moreover, the limit of $R \to \infty$ is considered in the soul of definition (57) and thus $[f]_N$ tends to a Hermitian matrix for real valued function $f \in \mathcal{S}(\mathbb{T}^{2d} \times \mathbb{R}^{2(n-d)})$.

As a definition $\star_N$ is defined by;

$$\int_{\mathbb{T}^{2d}} f_1 \star_N \cdots \star_N f_k := \lim_{R \to \infty} \xi_N tr\{[f_1]_N \cdots [f_k]_N\}$$

(59)

with the factor of $\xi_N = (\frac{2\pi R}{2m+1})^{2d}$, and for any $k \in \mathbb{N}$. In appendix A, it is shown that (59) makes sense and thus $\star_N$ is defined definitely and independently from $k$. Also in appendix A, it is discussed that the condition of $R \to \infty$ kills all the terms which depend on the momenta $k_{\vec{q}}$ s, and keeps only the momenta independent terms in (57). This causes the factor of $(2m+1)^{2d}$ to be appeared in $\xi_N$. It can be easily seen that $\star_N$ tends to the translation-invariant product $\star$ defined by $\alpha$ as $N(m) \to \infty$. Also by (59) it is obvious that;

$$\int_{\mathbb{T}^{2d}} f_1 \star_N \cdots \star_N f_{k-1} \star_N f_k = \int_{\mathbb{T}^{2d}} f_k \star_N f_1 \star_N \cdots \star_N f_{k-1}.$$

(60)

Moreover, (59) implies that (appendix A);

$$f \star_N g = \pi \left( \sum_{p=0}^{m} \frac{1}{p!} \left( \sum_{i=1}^{m} \alpha_L(\vec{\partial})^i \otimes \alpha_R(\vec{\partial})^i \right)^p (f \otimes g) \right) + \cdots$$

(61)

for derivation operator $\vec{\partial} = (\partial_1, \ldots, \partial_{2d})$ and for the ordinary product map $\pi$. Note that the term of ... vanishes rapidly as; $N(m) \to \infty$. Actually $\star_N$ can be considered as a truncated form of $\star$ in order $m$. Indeed, (61) produces a generalized formula for $\theta$-expansion of Groenewold-Moyal product in matrix formulation. This expansion formula of translation-invariant noncommutative products provides a generalized version of Seiberg-Witten map [40] for translation-invariant noncommutative fields in the setting of matrix calculus.

Finally, it can also be checked that the Haar measure in (58) leads to;

$$\lim_{R\to\infty} (\frac{2\pi R}{2m+1})^{2d} \, tr[f]_N = \int_{\mathbb{T}^{2d}} f$$

(62)

for any smooth function $f$.

Note that since $\star_N$ is not precisely associative (appendix A), then the matricial formulation of (57), is not exactly a MQM. Therefore, it has been proven that, for any translation-invariant product $\star$, there exists a MQM as a large $N$ limit of a series of matricial formulations (57). Thus, in the context, the matrix formulation of (57), is roughly referred to as MQM.

## 4. Translation-Invariant Deformation Quantization

In this section, translation-invariant quantization structures are discussed in details in order to obtain a consistent framework to classify and study the matrix modeling quantizations of $C^\infty(M)$. This leads to a deep understanding of quantization of TNG theories and their intrinsically noncommutative effects such as non-locality and UV/IR mixing. The Morita equivalence [8] for algebras $\mathcal{A}$ and $M_N(\mathcal{A})$, $N \in \mathbb{N}$, reduces this investigation only to translation-invariant Abelian gauge theories.

### 4.1. $\alpha$-Cohomology for Translation-Invariant Products

In this subsection it is shown that the definition (43) provides a concrete framework to classify translation-invariant quantization structures in the setting of a cohomology theory. Essentially this classification leads to algebra isomorphism classes due to quantization structures on $C^\infty(\mathbb{R}^m)$ modulo commutative products.

Let $C^n(\mathbb{R}^m) \subseteq C^\infty(\mathbb{R}^m \times ... \times \mathbb{R}^m)$, $n \geq 1$, ($n$ copies of $\mathbb{R}^m$) be complex vector spaces generated by smooth functions $f$ with $f(p, q, ..., q) = 0$ ($n \geq 3$), $f(p,p) = f(p,0) = 0$ ($n = 2$) and $f(0) = 0$ ($n = 1$), $p, q \in \mathbb{R}^m$. Then, consider the linear maps

$$\partial := \partial_n : C^n(\mathbb{R}^m) \to C^{n+1}(\mathbb{R}^m) \tag{63}$$

defined by;

$$\partial_n \sigma(p_0, ..., p_n) := \varepsilon_n \sum_{i=0}^{n} (-)^i \sigma(p_0, ..., p_{i-1}, \hat{p}_i, p_{i+1}, ..., p_n)$$

$$+ \varepsilon_n (-)^{n+1} \sigma(p_0 - p_n, ..., p_{n-1} - p_n), \tag{64}$$

$\sigma \in C^n(\mathbb{R}^m)$, with $\varepsilon_n = 1$ for odd $n$ and $\varepsilon_n = i$ for $n$ even. It is seen that $C^n(\mathbb{R}^m)$s as cochains and $\partial_n$s as coboundary maps define a complex with;

$$0 = C^0(\mathbb{R}^m) \xrightarrow{\partial_0} C^1(\mathbb{R}^m) \xrightarrow{\partial_1} ... \xrightarrow{\partial_{n-1}} C^n(\mathbb{R}^m) \xrightarrow{\partial_n} ... . \tag{65}$$

From (65) it can be seen that $\partial^2 = \partial_n \circ \partial_{n-1} = 0$ for any $n \in \mathbb{N}$. The complex (65) defines a cohomology theory so called $\alpha$-cohomology [61] which is similar to Hochschild one in definition [7]. Conventionally we use the notation of $\alpha_1 \sim \alpha_2$ for two $\alpha$-cohomologous $n$-cocycles $\alpha_1$ and $\alpha_2$. Also the cohomology class of $\alpha \in C^n(\mathbb{R}^m)$ is shown by $[\alpha]$. Therefore, the $\alpha$-cohomolgy group, $H_\alpha^n(\mathbb{R}^m) := Ker\partial_n / Im\partial_{n-1}$, classifies $n$-cocycles differing in coboundary terms into the same equivalence classes. Now consider the translation-invariant products given by $\alpha \in C^\infty(\mathbb{R}^m \times \mathbb{R}^m)$ due to definition (43). According to (44), associativity of $\star$ is equivalent to $\partial \alpha = 0$. Indeed, $H_\alpha^2(\mathbb{R}^m)$ classifies all the translation-invariant quantization structures over $C^\infty(\mathbb{R}^m)$ modulo coboundary terms. It can be easily seen from (52) that if $[\alpha] = 0$ then $\alpha$ is commutative. In the following it is shown that the inverse is also true and thus $H^2_\alpha(\mathbb{R}^m)$ classifies translation-invariant star products on $C^\infty(\mathbb{R}^m)$ modulo commutative ones. To see this fact set;

$$\alpha'(p,q) := \frac{1}{2}\left(\alpha(p,q) + \alpha(-p,-q)\right),$$

(66)

for any $p,q \in \mathbb{R}^m$. It can be checked that $\partial \alpha' = 0$ and thus it defines a translation-invariant structure on $C^\infty(\mathbb{R}^m)$. Next define $\star'$ with $\alpha'$ accordingly;

$$(f \star' g)(x) := \int \frac{\mathrm{d}^m p}{(2\pi)^m} \frac{\mathrm{d}^m q}{(2\pi)^m} \, \tilde{f}(q)\tilde{g}(p-q) e^{\alpha'(p,q)} e^{ip.x}.$$

(67)

for $f, g \in \mathcal{S}(\mathbb{R}^m)$. Now let $\alpha'' := \alpha - \alpha'$. More precisely;

$$\alpha''(p,q) = \frac{1}{2}\left(\alpha(p,q) - \alpha(-p,-q)\right).$$

(68)

for any $p, q \in \mathbb{R}^m$. Therefore, $\partial \alpha'' = 0$. Using (51) it can be shown that $\alpha''$ is commutative. Thus, $\alpha$ and $\alpha'$ differ in an associative commutative product. On the other hand; $\alpha''(p,q) = -\alpha''(-p,-q)$ for any $p,q \in \mathbb{R}^m$. Then, (46) and (49) lead to;

$$\alpha''(p,q) = -\alpha''(q,p)$$

(69)

for any $p, q \in \mathbb{R}^m$. Indeed $\alpha''$ obeys the following properties;

$$\begin{cases} \alpha''(p,q) = \alpha''(p, p-q) \\ \alpha''(p,q) = -\alpha''(-p,-q) \\ \alpha''(p,q) = -\alpha''(q,p) \end{cases},$$

(70)

for any $p, q \in \mathbb{R}^m$. In appendix B it is shown that (70) and $\partial \alpha'' = 0$ give hand an element of $C^1(\mathbb{R}^m)$, say $\beta$, which $\alpha'' = \partial \beta$. In fact, $\alpha''$ is a coboundary and thus; $\alpha \sim \alpha'$.

Now we are ready to show that $\alpha$ is commutative if and only if $\alpha$ is a coboundary. As mentioned above the only if part is obvious, so it is sufficient to prove the if term. According to appendix B, to prove this fact, it is enough to show that $\alpha'$ is also a coboundary provided $\alpha$ is commutative. Using (51) for commutative $\alpha$, (66) leads to;

$$\alpha'(p,q) = \frac{1}{2}(\alpha(0,q) - \alpha(0,p) + \alpha(0,p-q)) = \frac{1}{2}\partial\alpha_0(p,q),$$

(71)

for $\alpha_0(.) = \alpha(0,.) \in C^1(\mathbb{R}^m)$ and for any $p, q \in \mathbb{R}^m$. This proves that $\alpha_1 \sim \alpha_2$ if and only if $\alpha_1 - \alpha_2$ is commutative.

In fact (66) corresponds to each 2-cocycle $\alpha$ an $\alpha$-cohomologous element $\alpha'$ with property

$$\alpha'(p,q) = \alpha'(-p,-q),$$

(72)

for any $p, q \in \mathbb{R}^m$. It can also be seen that

$$\alpha(p,q) = \alpha_-(p,q) + \alpha_+(p,q)$$

(73)

with

$$\alpha_-(p,q) := \frac{1}{2}(\alpha(p,q) - \alpha(-p,q-p)),$$

$$\alpha_+(p,q) := \frac{1}{2}(\alpha(p,q) + \alpha(-p,q-p)),$$

(74)

defines another such correspondence for 2-cocycle $\alpha$. By (51) it can be seen that $\alpha_+ = \frac{1}{2}\partial\alpha_0$ and thus $\alpha$ and $\alpha_-$ define two $\alpha$-cohomologous 2-cocycles. In fact, $\alpha_-$ satisfies the condition

$$\alpha_-(p, q) = -\alpha_-(-p, q - p)$$

(75)

for any $p, q \in \mathbb{R}^m$. Therefore, according to (66) and (73), one can correspond to each 2-cocycle $\alpha$ an $\alpha$-cohomologous element $\alpha'_- := \alpha'_- = \alpha_-'$ with

$$\alpha'_-(p, q) = \alpha'_-(-p, -q) = -\alpha'_-(-p, q - p).$$

(76)

It is claimed that such correspondence is unique. That is for any cohomology class $[\alpha] \in H_\alpha^2(\mathbb{R}^m)$ there is a unique element $\alpha'_- \in [\alpha]$ satisfying (76). This can be considered as the Hodge theorem [96] for $\alpha$-cohomology classes of $H_\alpha^2(\mathbb{R}^m)$. Indeed $\alpha'_-$ is the unique solution in $[\alpha] \in H_\alpha^2(\mathbb{R}^m)$ for the equation

$$\Delta\alpha(p, q) := \alpha(0, q) - \alpha(0, p) + \alpha(0, p - q) + \alpha(p, q) + \alpha(p, p - q) = 0$$

(77)

$p, q \in \mathbb{R}^m$. Here $\Delta$ can be considered as a Laplace-Beltrami operator on the cochain $C^2(\mathbb{R}^m)$ of complex (65). Therefore, $\alpha'_-$ is called the harmonic element or the harmonic form of $[\alpha]$. Similarly a harmonic translation-invariant product is a translation-invariant product defined by a harmonic form. A collection of complicated manipulations shows that

$$\alpha'_-(p, q) = \frac{1}{2}(\alpha(p + q, q) - \alpha(p + q, p)),$$

(78)

and

$$\alpha'_-(p + nq, q) = \alpha'_-(p, q),$$

(79)

for any $p, q \in \mathbb{R}^m$ and for any $n \in \mathbb{Z}$. Equations (78) and (79) will be proven in the next subsection. To prove the Hodge theorem for $\alpha$-cohomology classes consider two harmonic $\alpha$-cohomologous 2-cocycles $\alpha_1$ and $\alpha_2$. It is enough to show that; $\alpha_1 = \alpha_2$. To see this let $\partial \beta = \alpha_1 - \alpha_2$. Thus for any $p, q \in \mathbb{R}^m$, (76) leads to;

$$\beta(-q) - \beta(-p) + \beta(q-p) = -(\beta(q-p) - \beta(-p) + \beta(-q)),$$

(80)

and hence; $\partial \beta = 0$. This proves the claim. According to (76), (49) and (51), any harmonic form $\alpha$ of an $\alpha$-cohomology class, obeys the properties of;

$$\begin{cases} \alpha(p,q) = -\alpha(p, p-q) \\ \alpha(p,q) = \alpha(-p, -q) \\ \alpha(p,q) = -\alpha(q,p) \end{cases},$$

(81)

for any $p, q \in \mathbb{R}^m$. The most important property of harmonic forms is cleared in integrating. In fact, it can be shown that for $\star$, a harmonic translation-invariant product, one finds that;

$$\int_{\mathbb{R}^m} f \star g = \int_{\mathbb{R}^m} fg$$

(82)

for any $f, g \in \mathcal{S}(\mathbb{R}^m)$. To see this note that by integration by part and by using the notation of subsection 3.2; $\alpha(p+q, p) = \langle \alpha_L(ip) | \alpha_R(iq) \rangle$, $p, q \in \mathbb{R}^{2n}$, one can show the following equality for any 2-cocycle $\alpha$;

$$\int \frac{d^m p}{(2\pi)^m} \frac{d^m p}{(2\pi)^m} d^m x \ \tilde{f}(p)\tilde{g}(q) \left(\alpha(p+q, p)\right)^k e^{i(p+q).x}$$

$$= \int \frac{d^m p}{(2\pi)^m} \frac{d^m p}{(2\pi)^m} d^m x \ \tilde{f}(p)\tilde{g}(q) \left(\alpha(p+q, p)\right)^{k-1} \alpha(-p-q, -q) \ e^{i(p+q).x},$$

(83)

for any $f, g \in \mathcal{S}(\mathbb{R}^m)$. This together with (81) leads to;

$$\int \frac{\mathrm{d}^m p}{(2\pi)^m} \frac{\mathrm{d}^m p}{(2\pi)^m} \mathrm{d}^m x \; \tilde{f}(p)\tilde{g}(q) \, (\alpha(p+q,p))^k e^{i(p+q).x} = 0 \,, \tag{84}$$

for harmonic form $\alpha$. This proves (82). As an example the Groenewold-Moyal product, $\star_{G-M}$, and the Wick-Voros production, $\star_{W-V}$, correspond to two $\alpha$-cohomologous 2-cocycles $\alpha_{G-M}$ and $\alpha_{W-V}$, respectively [61]. In fact, $\alpha_{G-M}$ is the harmonic form of the $\alpha$-cohomology class $[\alpha_{G-M}]$ and thus $\star_{G-M}$ satisfies the condition (82).

The other prominent property of harmonic forms is cleared in the setting of $\alpha^*$-cohomology theory. $\alpha^*$-cohomology is defined by complex;

$$0 = C_*^0(\mathbb{R}^m) \xrightarrow{\partial_0} C_*^1(\mathbb{R}^m) \xrightarrow{\partial_1} \dots \xrightarrow{\partial_{n-1}} C_*^n(\mathbb{R}^m) \xrightarrow{\partial_n} \dots \,, \tag{85}$$

where $C_*^n(\mathbb{R}^m)$ is the cochain of elements $f \in C^n(\mathbb{R}^m)$, with property

$$f^*(p_1, \dots, p_n) = f(-p_1, p_n - p_1, p_{n-1} - p_1, \dots, p_2 - p_1) \,, \tag{86}$$

for $f^*$ the complex conjugate of $f$ and for any collection of $p_1, \dots, p_n \in \mathbb{R}^m$. Conventionally $H_{\alpha^*}^n(\mathbb{R}^m)$ is used for the $n$th cohomology group of complex (85), or more precisely for the $n$th $\alpha^*$-cohomology group. Moreover, the complex inclusion map

$$\begin{array}{ccccccccc}
0 = C_*^0(\mathbb{R}^m) & \xrightarrow{\partial_0} & C_*^1(\mathbb{R}^m) & \xrightarrow{\partial_1} & \dots & \xrightarrow{\partial_{n-1}} & C_*^n(\mathbb{R}^m) & \xrightarrow{\partial_n} & \dots \\
& & \downarrow i_0 & & \downarrow i_1 & \dots & & \downarrow i_n & \\
0 = C^0(\mathbb{R}^m) & \xrightarrow{\partial_0} & C^1(\mathbb{R}^m) & \xrightarrow{\partial_1} & \dots & \xrightarrow{\partial_{n-1}} & C^n(\mathbb{R}^m) & \xrightarrow{\partial_n} & \dots
\end{array} \tag{87}$$

leads to a family of inclusions of cohomology groups;

$$i_{n_*}: H^n_{\alpha^*}(\mathbb{R}^m) \hookrightarrow H^n_\alpha(\mathbb{R}^m) \,.$$

(88)

Indeed any 2-cocycle $\alpha \in C^2_*(\mathbb{R}^m)$ defines a translation-invariant product $\star$ over $C^\infty(\mathbb{R}^m)$ with property

$$(f \star g)^* = g^* \star f^*,$$

(89)

for any $f, g \in \mathcal{S}(\mathbb{R}^m)$. Such productions are usually called involutive products. In fact the 2-cocycles of involutive translation-invariant products satisfy the condition

$$\alpha^*(p, q) = \alpha(-p, q - p)\,,$$

(90)

for any $p, q \in \mathbb{R}^m$. Thus $H^2_{\alpha^*}(\mathbb{R}^m)$ classifies involutive translation-invariant deformations of $C^\infty(\mathbb{R}^m)$ up to commutativity. Moreover since; $\partial f^* = \pm(\partial f)^*$, and $f^* \in C^n_*(\mathbb{R}^m)$, for any $f \in C^n_*(\mathbb{R}^m)$, then $\alpha^* \in C^2_*(\mathbb{R}^m)$ is a 2-cocycle when $\alpha$ is a 2-cocycle. Thus to any $\alpha^*$-cohomology class $[\alpha] \in H^2_{\alpha^*}(\mathbb{R}^m)$, one can naturally correspond a conjugate $\alpha^*$-cohomology class by $[\alpha^*]$. The condition (90) together with (51) asserts that if $[\alpha]$ belongs to $H^2_{\alpha^*}(\mathbb{R}^m) \subseteq H^2_\alpha(\mathbb{R}^m)$, then $[\alpha^*]$ is the dual of $[\alpha]$ in the sense of;

$$[\alpha] + [\alpha^*] = 0\,.$$

(91)

This is the pure imaginary condition for $\alpha$-cohomology classes. Moreover, it can be seen that the Hodge theorem is true for $H^2_{\alpha^*}(\mathbb{R}^m)$ and thus any harmonic form $\alpha$ of an $\alpha^*$-cohomology class of $H^2_{\alpha^*}(\mathbb{R}^m)$, is pure imaginary;

$$\alpha^* = -\alpha\,.$$

(92)

Indeed, it can be seen that $\alpha'^{*}_{-}(p,q) = \alpha'_{-}(-p, q-p)$ for any $\alpha \in C^2_*(\mathbb{R}^m)$ and for any $p,q \in \mathbb{R}^m$. Thus the Hodge theorem is also true for $\alpha^*$-cohomology and therefore, (92) follows, since for $\alpha$ a 2-cocycle of $C^2_*(\mathbb{R}^m)$, one finds that; $\alpha'^{*}_{-} = -\alpha'_{-}$. Conversely (81) says that $\alpha$ is a harmonic form if and only if $\alpha^*$ is harmonic. Therefore, uniqueness of harmonic forms asserts that $H^2_{\alpha^*}(\mathbb{R}^m)$ is the collection of all pure imaginary elements of $H^2_\alpha(\mathbb{R}^m)$.

Finally it can be easily seen that, the harmonic form $\alpha$ of a pure imaginary $\alpha$-cohomology class, satisfies the condition

$$\alpha^*(p,q) = \alpha(p, p-q),$$

(93)

for any $p,q \in \mathbb{R}^m$, which can be compared with commutativity condition (52).

### 4.2. Structures of Translation-Invariant Products

In this subsection, translation-invariant products are discussed in the setting of loop calculations for field theories with translation-invariant products. This leads to studying the translation-invariant structures more exhaustively.

It would be interesting to study the role of $\alpha$-cohomology in classifying the algebraic structures on $C^\infty(\mathbb{R}^m)$ due to translation-invariant products. By definition two translation-invariant products $\star_1$ and $\star_2$ are equivalent if there exists an invertible translation-invariant linear map $T: C^\infty(\mathbb{R}^m) \to C^\infty(\mathbb{R}^m)$, such that;

$$T(f \star_1 g) = T(f) \star_2 T(g)$$

(94)

for any $f, g \in \mathcal{S}(\mathbb{R}^m)$. Strictly speaking $T$ is an algebra isomorphism from $\mathcal{S}(\mathbb{R}^m)_{\star_1}$ to $\mathcal{S}(\mathbb{R}^m)_{\star_2}$ with $T(\partial_\mu f) = \partial_\mu T(f)$ for any $f \in \mathcal{S}(\mathbb{R}^m)$ and $1 \leq \mu \leq m$. Conventionally, the notation of $\star_1 \sim \star_2$ is used for two equivalent translation-invariant products $\star_1$ and $\star_2$. It can be shown that $\star_1 \sim \star_2$ if and only if $\alpha_1 \sim \alpha_2$. To see this, according to (94) assume that $\star_1 \sim \star_2$ with $T = e^\Delta$ for $\Delta$ a translation-invariant linear differential operator. Thus;

$$\widetilde{T(f)}(p) = \int \mathrm{d}^m x \, e^{-ip.x} e^\Delta f = e^{\widetilde{\Delta}(p)} \tilde{f}(p),$$

(95)

for any $p, q \in \mathbb{R}^m$ and for any $f \in \mathcal{S}(\mathbb{R}^m)$. Note that since $T$ is invertible then, $T(1) = 1$. Thus $\widetilde{\Delta}(0) = 0$ and hence; $\widetilde{\Delta} \in C^1(\mathbb{R}^m)$. Therefore;

$$(T(f) \star_2 T(g))(x)$$

$$= \int \frac{\mathrm{d}^m p}{(2\pi)^m} \frac{\mathrm{d}^m p}{(2\pi)^m} \, \widetilde{T(f)}(q) \, \widetilde{T(g)}(p-q) \, e^{\alpha_2(p,q)} \, e^{ip.x}$$

$$= \int \frac{\mathrm{d}^m p}{(2\pi)^m} \frac{\mathrm{d}^m p}{(2\pi)^m} \, \tilde{f}(q) \, \tilde{g}(p-q) \, e^{\alpha_2(p,q)+\widetilde{\Delta}(q)+\widetilde{\Delta}(p-q)} \, e^{ip.x}$$

$$= \int \frac{\mathrm{d}^m p}{(2\pi)^m} \frac{\mathrm{d}^m p}{(2\pi)^m} \, \tilde{f}(q) \, \tilde{g}(p-q) \, e^{\alpha_2(p,q)+\partial\widetilde{\Delta}(p,q)+\widetilde{\Delta}(p)} \, e^{ip.x}$$

$$= \int \frac{\mathrm{d}^m p}{(2\pi)^m} \frac{\mathrm{d}^m p}{(2\pi)^m} \, \widetilde{T(f \star g)}(p) \, e^{ip.x}$$

$$= T(f \star g)(x),$$

(96)

for any $f, g \in \mathcal{S}(\mathbb{R}^m)$ and for $\star$ the translation-invariant product defined by $\alpha_2 + \partial\widetilde{\Delta}$. Thus (94) asserts that; $\alpha_1 = \alpha_2 + \partial\widetilde{\Delta}$, and consequently $\alpha_1 \sim \alpha_2$. Now conversely assume that $\alpha_1 = \alpha_2 + \partial\widetilde{\Delta}$ for $\widetilde{\Delta} \in C^1(\mathbb{R}^m)$. By (94) and (96) it is enough to define; $T := e^\Delta$.

It can be seen that the $\alpha$-cohomology class of a translation-invariant product $\star$, solely describes the UV/IR mixing behavior of translation-invariant $\phi^4_\star$ theory [61]. In fact, the non-planar corrections to 2-point functions in $\phi^4_\star$ theory on $\mathbb{R}^4$ at one-loop level are given by;

$$G^{(2)}_{NP/1}(p) := \int \frac{d^4q}{(2\pi)^4} \frac{e^{-\alpha(0,p)+\omega(p,q)}}{(p^2-m^2)^2(q^2-m^2)},$$

(97)

with

$$\omega(p,q) = \omega_\alpha(p,q) := \alpha(p+q,p) - \alpha(p+q,q)$$

(98)

for any $p, q \in \mathbb{R}^4$. It is easily seen by (52) that $\omega_\alpha = 0$ if and only if $\alpha$ is commutative. This lets one to define $\omega$ on the elements of $H^2_\alpha(\mathbb{R}^m)$. Actually, the restriction of $\omega$ to $H^2_{\alpha^*}(\mathbb{R}^m)$, classifies the UV/IR mixing of involutive translation-invariant $\phi^4_\star$ theory at one loop corrections. Moreover, it is known [97] that such classification comes true in all order of perturbation.

We claime that this classification of UV/IR mixing with $\alpha$-cohomology is true for all possible translation-invariant (noncommutative) field theories. To see this, consider two noncommutative versions of a given quantum field theory due to two equivalent translation-invariant (noncommutative) products $\star_1$ and $\star_2$. Assume that the theory contains $n$ fields $\{\phi_i\}_{i=1}^n$. Consider the most general interaction term in the action;

$$S^{int}_\star = \int_{\mathbb{R}^m} \phi_{i_1,(\mu_{i_1})} \star \ldots \star \phi_{i_k,(\mu_{i_k})},$$

(99)

where $\phi_{,(\mu)} = \frac{\partial^p \phi}{\partial x^{\mu_1} \ldots \partial x^{\mu_p}}$ for multi-plet $(\mu) = (\mu_1, \ldots, \mu_p)$. $\star$ in (98) stands for $\star_1$ and $\star_2$, respectively. For the case of $\star = \star_1$, the interaction term (99) can be rewritten in the phase space with;

$$S^{int}_{\star_1} = \int \prod_{j=1}^k \left(\frac{d^m p^j}{(2\pi)^m}\right) p^1_{(\mu_{i_1})} \widetilde{\phi_{i_1}}(p^1) \ldots p^k_{(\mu_{i_1})} \widetilde{\phi_{i_k}}(p^k) V^k_{\star_1}(p^1, \ldots, p^k) \delta^{(m)}\left(\sum_{j=1}^k p^j\right)$$

(100)

for

$$V_{\star_1}^k(p^1, \ldots, p^k) = e^{\sum_{i=2}^k \alpha_1(\sum_{j=1}^i p^j, \sum_{j=1}^{i-1} p^j)}$$

(101)

the noncommutative vertex. According to (95), the redefinition of $\phi_i' = e^{\Delta}\phi_i$, $i = 1, \ldots, n$, gives $S_{\star_1}^{int}$ in terms of $V_{\star_2}^k$ with;

$$S_{\star_1}^{int} = \int \prod_{j=1}^k \left(\frac{d^m p^j}{(2\pi)^m}\right) p_{(\mu_{i_1})}^1 \widetilde{\phi_{i_1}'}(p^1) \ldots p_{(\mu_{i_1})}^k \widetilde{\phi_{i_1}'}(p^k) V_{\star_2}^k(p^1, \ldots, p^k) \delta^{(m)}\left(\sum_{j=1}^k p^j\right).$$

(102)

More precisely

$$S_{\star_1}^{int}(\phi_1, \ldots, \phi_n) = S_{\star_2}^{int}(\phi_1', \ldots, \phi_n').$$

(103)

Thus, it is enough to show that the propagators of $\phi_i'$s, $i = 1, \ldots, n$, are also given in terms of $\alpha_2$. The propagators $\tilde{G}_{\star_1}(\phi_i, \phi_j)$, $i, j = 1, \ldots, n$, are given by

$$\tilde{\Xi}_{\star_1}^{ij}(p) \ \tilde{G}_{\star_1}(\phi_i, \phi_j)(p) = 1$$

(104)

for

$$S_{\star_1}^0 = \sum_{i,j=1}^n \int \frac{d^m p}{(2\pi)^m} \widetilde{\phi_i}(p)\widetilde{\phi_j}(-p)\tilde{\Xi}_{\star_1}^{ij}(p).$$

(105)

Moreover it can be seen that;

$$\tilde{\Xi}_{\star_1}^{ij}(p) = \tilde{\Xi}^{ij}(p) \, e^{\alpha_1(0,p)},$$

(106)

for any $p \in \mathbb{R}^m$ and for $\widetilde{\Xi}^{ij}$ the Fourier transformed form of differential operator $\Xi^{ij}$;

$$S^0 = \sum_{i,j=1}^{n} \int_{\mathbb{R}^m} \phi_i \Xi^{ij} \phi_j \,.$$

(107)

It can be seen that replacing $\phi_i$ by $\phi'_i$ leads to

$$S^0_{\star_1} = \sum_{i,j=1}^{n} \int \frac{d^m p}{(2\pi)^m} \widetilde{\phi'_i}(p) \widetilde{\phi'_j}(-p) \Xi^{ij}_{\star_2}(p) \,.$$

(108)

Thus,

$$S^0_{\star_1}(\phi_1, \ldots, \phi_n) = S^0_{\star_2}(\phi'_1, \ldots, \phi'_n) \,,$$

(109)

and hence;

$$\widetilde{G}_{\star_1}(\phi_i, \phi_j) = \widetilde{G}_{\star_2}(\phi'_i, \phi'_j)$$

(110)

for $i,j = 1, \ldots n$. Therefore, quantum corrections and loop calculations for two translation-invariant field theories defined by equivalent star products $\star_1$ and $\star_2$ lead to the same results. More precisely moving through an $\alpha$-cohomology class of $H^2_\alpha(\mathbb{R}^m)$, produces no new physics. For example Wick-Voros noncommutative field theories, have no new quantum behaviors in compare with Groenewold-Moyal ones. Therefore, it seems that all the abnormal effects of Wick-Voros noncommutative field theories such as UV/IR mixing, non-locality and consequently (non-) renormalizability, coincide with those of Groenewold-Moyal ones. Consequently, it would be expected that the Grosse-Wulkenhaar approach [36, 37] and the method of $1/p^2$ [38] also work for renormalizing Wick-Voros $\phi^4_{\star_{V-W}}$ theory [97].

As it was stated above, the quantum corrections for a translation-invariant $\phi^4_\star$ theory are described by $\omega$ as a character of the $\alpha$-cohomology class of $\star$ [97]. In fact, for translation-invariant product $\star$, $\omega$ can be equivalently computed for the harmonic form of $\alpha$-cohomology class of $\star$. Indeed, (98) yields;

$$\omega(p,q) = \omega_{\alpha_H}(p,q) = 2\alpha_H(p+q,p),$$

(111)

for $\alpha_H$, the harmonic form of $\alpha$-cohomology class of $\star$. More precisely;

$$\alpha_H(p,q) = -\frac{1}{2}\omega(p-q,q).$$

(112)

This with uniqueness of harmonic forms implies that;

$$\alpha'_-(p,q) = -\frac{1}{2}\omega_\alpha(p-q,q) \ .$$

(113)

for any 2-cocycle $\alpha$, which could be seen by (98) and setting $\alpha = \alpha'_- + \partial\beta$ for a coboundary $\partial\beta$. So one finds that; $\omega_\alpha(p,q) = 2\alpha'_-(p+q,p)$. It can be precisely shown that the quantum corrections for any translation-invariant field theory are described by $\omega$ in all orders. To see this, at the first step it must be shown that the translation-invariant products described by coboundaries affect the Feynman diagrams amplitudes only in terms of external momenta. Indeed, it can be seen by induction that;

$$\sum_{i=2}^{n}\partial\beta(\sum_{j=1}^{i}p^j, \sum_{j=1}^{i-1}p^j) = \sum_{i=1}^{n}\beta(p^i) - \beta(\sum_{i=1}^{n}p^i)$$

(114)

for any $\beta \in C^1(\mathbb{R}^m)$ and for any collection of $p^1, \ldots, p^n \in \mathbb{R}^m$. Therefore, by momentum conservation law at vertices, the noncommutative vertex (101) for a translation-invariant product $\star$ defined by $\partial\beta$ is

$$V_\star^k(p^1, \ldots, p^k) = e^{\sum_{i=1}^k \beta(p^i)}.$$

(115)

On the other hand, the phase factor of propagators is;

$$e^{-\partial \beta(0,p)} = e^{-\beta(p) - \beta(-p)}$$

(116)

$p \in \mathbb{R}^m$. This cancels the relevant phase factors of initial and final vertices. Therefore, (116) together with (115) cancels out all the internal momentum dependent phase factors and keeps only the phase factors of external momenta.

The results above, lets one to study the role of star products in loop calculations only for harmonic forms. Since for harmonic form $\alpha_H$, we have $\alpha_H(0,p) = 0$, $p \in \mathbb{R}^m$, then there is no nontrivial phase factor for the propagators. Moreover, by (101) and (111) the noncommutative vertex is

$$V_\star^k(p^1, \ldots, p^k) = e^{\sum_{i=1}^{k-1} \frac{1}{2} \omega(\sum_{j=1}^i p^j, p^{i+1})}.$$

(117)

This shows that the quantum corrections of a translation-invariant field theory not are entirely related to $\alpha$-cohomology class of its star product, but they are precisely described by $\omega$ as a character of $\alpha$-cohomology classes.

Using the properties of harmonic forms in (81) one can show that;

$$\begin{cases} \omega(p,q) = -\omega(q,p) \\ \omega(p,q) = \omega(-p,-q), \\ \omega(p,0) = 0 \end{cases}$$

(118)

for any $p, q \in \mathbb{R}^m$.

On the other hand, (44) asserts that (98) can be written in the form of;

$$\omega(p,q) = \alpha(p, p-q) - \alpha(p,q) = \omega(p-q, q),$$

(119)

and consequently;

$$\omega(p,q) = \omega(p + nq, q)$$

(120)

for any $n \in \mathbb{Z}$. The property (119) together with (112) leads to

$$\alpha_H(p,q) = -\frac{1}{2}\omega(p,q)$$

(121)

and then proves (78) by (98) and (113). Thus, $\omega$ satisfies the associativity condition of (44). This obviously confirms the cohomological description of loop calculations in translation-invariant field theories. On the other hand, again by uniqueness of harmonic elements, (120) and (112) lead to (79). Moreover, by (118) and (120) it is seen that;

$$\omega(p,q) = -\omega(p,-q),$$

(122)

and thus; $\alpha_H(p,q) = -\alpha_H(p,-q)$. Moreover, in appendix C it is shown that, (120) leads to

$$\omega(rp, p) = 0,$$

(123)

for any $r \in \mathbb{Q}$, and thus by continuity of $\omega$;

$$\omega(rp, p) = 0,$$

(124)

for any $r \in \mathbb{R}$.

It may seem that conditions (118)-(124) would lead to $\omega$ or equivalently $\alpha_H$, being an anti-symmetric bilinear form on $\mathbb{R}^m$, but it can be shown that $\omega$ can take more general forms. For example

$$\widetilde{\omega}(p,q) = p^\mu \theta^0_{\mu\nu} q^\nu + p^\mu \theta^1_{\mu\nu} q^\nu p^\sigma \theta^2_{\sigma\lambda} q^\lambda p^\zeta \theta^3_{\zeta\xi} q^\xi ,$$

(125)

$p, q \in \mathbb{R}^m$, for antisymmetric matrices $\theta^i$, $i = 0,1,2,3$, satisfies all the conditions (118)-(124). Moreover, it can be seen that $\partial \widetilde{\omega} = 0$ and thus (125) defines a translation-invariant product according to (121). On the other hand, since $\omega$ in (125) is a harmonic form, its star product differs from Groenewold-Moyal one in $\alpha$-cohomology class for nonzero $\theta^i$, $i = 1,2,3$.

It would be also an interesting issue to study the relation between noncommutativity of space and abnormal quantum behaviors of translation-invariant noncommutative field theories. As it was mentioned above, this investigation leads to studying the relation of $\omega$ and $[f, g]_\star$ for any Schwartz functions $f$ and $g$. Indeed, it is the question that how much of noncommutativity of space is reflected by the $\alpha$-cohomology class of the star product. To answer this question, one may focus on commutation relations of Fourier modes. Consider a 2-cocycle $\alpha$ and its star product $\star$. Note that;

$$(e^{ip.x} \star e^{iq.x})(x)$$
$$= \int \frac{d^m k_1}{(2\pi)^m} \frac{d^m k_2}{(2\pi)^m} d^m y d^m z \; e^{ip.y} e^{iq.z} e^{-ik_1.y} e^{-ik_2.z} e^{\alpha(k_1+k_2, k_1)} e^{i(k_1+k_2).x}$$
$$= e^{\alpha(p+q,p)} e^{i(p+q).x} = e^{\alpha_H(p+q,p)} e^{\partial\beta(p+q,p)} e^{i(p+q).x} ,$$

(126)

with $\alpha = \alpha_H + \partial\beta$ for harmonic form $\alpha_H \in [\alpha]$, and coboundary $\partial\beta$. Therefore;

$$[e^{ip.x}, e^{iq.x}]_\star$$
$$= \left( e^{\alpha_H(p+q,p)} - e^{-\alpha_H(p+q,p)} \right) e^{\partial\beta(p+q,k_1)} e^{i(p+q).x} .$$

(127)

Thus, for two equivalent 2-cocycles $\alpha_1$ and $\alpha_2$ with $\alpha_1 = \alpha_2 + \partial\beta$ one finds;

$$[e^{ip.x}, e^{iq.x}]_{\star_1} = e^{\partial\beta(p+q,p)} [e^{ip.x}, e^{iq.x}]_{\star_2}.$$

(128)

Therefore, $\alpha$-cohomology classes determine noncommutativity of space up to phase factors. Conversely, it can be seen that if star products $\star_1$ and $\star_2$ satisfy (128) for all Fourier modes and for a commutative 2-cocycle $\partial\beta$, then $\star_1 \sim \star_2$. To see this note that by (127) and (128) one has;

$$\left(e^{\alpha_H^1(p+q,p)} - e^{-\alpha_H^1(p+q,p)}\right) e^{\partial\beta^1(p+q,p)}$$
$$= \left(e^{\alpha_H^2(p+q,p)} - e^{-\alpha_H^2(p+q,p)}\right) e^{\partial\beta(p+q,p)} e^{\partial\beta^2(p+q,p)},$$

(129)

where $\alpha_i = \alpha_H^i + \partial\beta^i$, $i = 1,2$, as mentioned above. Then

$$e^{\alpha_H^1(p+q,p)} - e^{-\alpha_H^1(p+q,p)} = \left(e^{\alpha_H^2(p+q,p)} - e^{-\alpha_H^2(p+q,p)}\right) e^{\partial\gamma(p+q,p)},$$

(130)

for $\gamma = \beta + \beta^2 - \beta^1$. By (81), $e^{\alpha_H^i(p+q,p)} - e^{-\alpha_H^i(p+q,p)}$, $i = 1,2$, are antisymmetric in exchanging $p \leftrightarrow p+q$, and thus;

$$\partial\gamma(p+q,p) = \partial\gamma(p,p+q)$$

(131)

for any $p, q \in \mathbb{R}^m$. Setting $p = 0$, one finds that;

$$\gamma(q) = -\gamma(-q).$$

(132)

Therefore, by (131) and (132) we have;

$$\partial\gamma = 0,$$

(133)

which by (130) and (79) leads to;

$$e^{\alpha_H^1(p,q)} - e^{-\alpha_H^1(p,q)} = e^{\alpha_H^2(p,q)} - e^{-\alpha_H^2(p,q)},$$

(134)

for any $p, q \in \mathbb{R}^m$. It can be easily seen that for two non-vanishing functions $f$ and $g$, the equality;

$$f - \frac{1}{f} = g - \frac{1}{g}$$

(135)

leads to either $f = g$ or $fg = -1$. But $e^{\alpha_H^1(p,q)} e^{\alpha_H^2(p,q)} \neq -1$ at least over an open set around $p = q = 0$. Therefore, $\alpha_H^1$ and $\alpha_H^2$ coincide over an open set around the origin. On the other hand, by assumption, $\alpha_H^1$ and $\alpha_H^2$, are entirely determined by their germs or their Taylor expansions at origin. Thus, we conclude that $\alpha_H^1 = \alpha_H^2$ and consequently; $\star_1 \sim \star_2$. This proves our claim.

## 5. Quantization and Consistent Anomalies of Matrix Modeled Gauge Theories

In this section, MMG theories are quantized in (anti-) BRST and path integral formulations. Moreover, the anomalous behaviors of TNG theories are calculated with non-perturbative methods coming from the abilities of matrix modeling formalism. Indeed, the matrix modeling approach translates the complicated loop calculations in TNG theories into a formulation of semi-commutative manipulations based on algebraic structures. These algebraic structures enable one to generalize the commutative concepts to their semi-commutative counterparts. This achievement helps one to extract the non-perturbative results for TNG theories without trapping in complicated loop calculations with noncommutative products and their relevant elaborations.

## 5.1. Geometric Quantization and Consistent Anomalies of Matrix Modeled Gauge Theories

In this subsection, we try to quantize MMG theories in the formalism of geometric quantization. It can be seen that this leads to a noncommutative version of (anti-) BRST transformations and consequently to (extended) BRST quantization of MMG theories. It is seen that this also produces a non-commutative counterpart for descent equations in the setting of geometric quantization. As it was mentioned above, MMG theories are asymptotically equivalent to TNG theories. Thus, it seems that the results of manipulations for a family of consistent $N \times N$-MMG theories lead to the result of the same calculations for the asymptotic TNG theory. Therefore, to calculate consistent anomalies and Schwinger terms for a given TNG theory, it can be equivalently apply the matrix modeling formulation. To do so, at the first step MMG theories should be quantized. For simplicity one should start with quantization of Abelian $N \times N$-MMG theories. Here it is followed the geometric quantization approach of [98, 99]. To proceed more strictly, a number of assumptions should be set initially. First, set $\mathcal{H}^2 = \mathcal{H} \otimes \mathcal{H}$ and assume that $\mathcal{H}^2$ is equipped with the ordinary induced algebraic structure. Suppose that $\mathcal{A}$, the set of Hermitian elements of $\mathcal{H}^2 \otimes \Omega^1_{\text{deR}/\mathcal{SP}}(M)$, is the Affine space of gauge fields. Moreover, consider that $\mathcal{G}$, the set of anti-Hermitian elements of $\mathcal{H}^2$, is the set of infinitesimal gauge transformations. Let $\mathcal{H}^2$ act on $\Gamma(S(M)_N)$ with;

$$\sum a_i \otimes b_i \triangleright \psi = \sum a_i \triangleright_\text{I} \psi \triangleleft S(b_i),$$

(136)

for $\sum a_i \otimes b_i \in \mathcal{H}^2$ and $\psi \in \Gamma(S(M)_N)$. It can be easily seen that these assumptions unify all the three types of actions (7)-(9). Specially, any element with action type I lives in the first component of $\mathcal{H}^2$ while the elements with action type II live in the second component of $\mathcal{H}^2$. Obviously, since any type III acting element can be expressed as a combination of two type I and type II acting elements, (136) unifies all types of actions. Indeed, (136) not only gives a unifying formalism, but it admits more general action types.

Now consider $\mathscr{g}$ as a Lie algebra with the ordinary Lie bracket and then formally correspond to it a Lie group $\mathcal{G}$ generated with formal elements of $e^\alpha = \sum_{k=0}^\infty \frac{\alpha^k}{k!}$, $\alpha \in \mathscr{g}$. Indeed, any element of $\mathcal{G}$ can be considered as a product of finitely many elements of $e^\alpha$. Also define $(e^\alpha)^{-1} = e^{-\alpha}$. $\mathcal{G}$ acts on $\mathcal{A}$ by;

$$A \cdot e^\alpha = -i \mathrm{d}_m \alpha + e^{-\alpha} A e^\alpha .$$

(137)

One should suppose that this action is free and thus (137) leads to a principal $\mathcal{G}$-bundle over the moduli space $\mathcal{A}/\mathcal{G}$;

$$\mathcal{G} \hookrightarrow \mathcal{A} \twoheadrightarrow \mathcal{A}/\mathcal{G} .$$

(138)

Set a principal connection over (138) with Cartan connection form $\Pi$, which sends the vertical elements of $T_A \mathcal{A}$ isomorphically to $\mathscr{g}$, for any $A \in \mathcal{A}$.

To proceed more, consider the semi-direct product group $\mathcal{G} \rtimes \mathcal{T}(M)$ with Lie algebra bracket $[X, \alpha] \coloneqq X(\alpha) \coloneqq \frac{d}{dt}\Big|_{t=0} \tau_{mX_t}(\alpha)$, for $X \in \text{Lie}\,\mathcal{T}(M)$ and $\alpha \in \mathscr{g}$. Next fix $A_0 \in \mathcal{H}^2 \otimes \Omega^1_{\text{deR}/\mathcal{SP}}(M)$ and define;

$$A = A_{(g,X_t)} \coloneqq -i g^{-1} \mathrm{d}_m g + g^{-1} \tau_{mX_t}(A_0) g \quad \in \mathcal{H}^2 \otimes \Omega^1_{\text{deR}}(\mathcal{G} \rtimes \mathcal{T}(M)) .$$

(139)

Then, define the ghost field with;

$$\omega = -i \Pi|_{\mathcal{A}_{A_0}} \quad \in \mathcal{H}^2 \otimes \Omega^1_{\text{deR}}(\mathcal{G} \rtimes \mathcal{T}(M)) ,$$

(140)

where $\Pi|_{\mathcal{A}_{A_0}}$ is the restriction of $\Pi$ to the fiber including $A_0$. Split the exterior derivative operator of $\mathcal{G} \rtimes \mathcal{T}(M)$, $d_{\mathcal{G} \rtimes \mathcal{T}(M)}$, into its cotangential components on $T^*\mathcal{T}(M)$ and $T^*\mathcal{G}$, respectively denoted by d and $\delta$. This leads to;

$$dA = d_m A \ , \ d\omega = -d_m \omega \ ,$$
$$\delta A = d\omega - iA\omega - i\omega A \ ,$$
$$\delta \omega = -i\omega^2 \ .$$

(141)

According to [98], $\delta$ is conventionally called the BRST transformation. Thus, (141) asymptotically leads to a stricter formulation of noncommutative BRST derivation in compare with [100, 101]. Indeed, in (141) the ghost field includes all types of actions (7)-(8) for infinitesimal gauge transformations in its substantial setting. On the other hand, $d^2 = d_{\mathcal{G} \rtimes \mathcal{T}(M)}{}^2 = 0$, implies that;

$$\delta^2 = \delta d + d\delta = 0 \ .$$

(142)

Following the axially extension method [99], one can work out the anti-BRST transformation and anti-ghost similarly. To do so, one should initially replace $\mathcal{H}^2$ (resp. $A_0$) with $\mathcal{H}^2 \oplus \mathcal{H}^2 \gamma_5$ (resp. $A_0 + B_0$, for $B_0$ the axial component of the gauge field). Next, $\mathcal{g}$ should be replaced by $\tilde{\mathcal{g}} := \mathcal{g} \oplus \mathcal{g}\gamma_5$, as the space of extended infinitesimal gauge transformations. The production of $\tilde{\mathcal{g}}$ is given by

$$(a + b\gamma_5).(c + d\gamma_5) = (ac + bd) + (ad + bc)\gamma_5$$

(143)

for $(a + b\gamma_5), (c + d\gamma_5) \in \mathcal{H}^2 \oplus \mathcal{H}^2 \gamma_5$. Subsequently, consider $\tilde{\mathcal{G}}$ as a Lie group generated by formal elements $e^\alpha = \sum_{k=0}^\infty \frac{\alpha^k}{k!}$, $\alpha \in \tilde{\mathcal{g}}$. By this notation ghost, $\omega$, and anti-ghost, $\omega^*$, are defined by;

$$\omega = P \circ -i\Pi|_{\mathcal{A}_{(A_0,B_0)}}, \omega^* = P_5 \circ -i\Pi|_{\mathcal{A}_{(A_0,B_0)}} \in (\mathcal{H}^2 \oplus \mathcal{H}^2 \gamma_5) \otimes \Omega^1_{\text{deR}}(\mathcal{G} \rtimes \mathcal{T}(M)) \ ,$$

(144)

for projections $P: \tilde{\mathscr{G}} \twoheadrightarrow \mathscr{g}$ and $P_5 = 1 - P$. Similarly split the exterior derivative operator of $\tilde{\mathcal{G}}$ into its components on $\mathscr{g}$ and $\mathscr{g}\gamma_5$ respectively denoted by $\delta$ and $\delta^*$. A direct calculation leads to the following results;

$$dA = d_m A \ , \ dB = d_m B \ , \ d\omega = -d_m \omega \ , \ d\omega^* = -d_m \omega^*,$$

$$\delta A = d\omega - iA\omega - i\omega A \quad , \quad \delta B = -iB\omega - i\omega B,$$

$$\delta\omega = -i\omega^2 \quad , \quad \delta\omega^* = -i\omega^*\omega - i\omega\omega^*,$$

$$\delta^* A = -iB\omega^* - i\omega^* B \quad , \quad \delta^* B = d\omega^* - iA\omega^* - i\omega^* A,$$

$$\delta^* \omega = 0 \quad , \quad \delta^* \omega^* = -i\omega^{*2}.$$

(145)

According to [99], $\delta$ and $\delta^*$ are respectively called BRST and anti-BRST derivation. Moreover, one can similarly show that;

$$\delta^2 = \delta^{*2} = \delta d + d\delta = \delta^* d + d\delta^* = \delta\delta^* + \delta^*\delta = 0.$$

(146)

The geometric quantization of a non-Abelian MMG theory can be taken place in the same way with replacing $\mathcal{H}^2$ by $U(\mathfrak{g}) \otimes \mathcal{H}^2$ in the given approach. The details of the elaborations are precisely similar to what was stated in section 2. Moreover the axially extension of non-Abelian MMG theories can be worked out by the same approach with replacing $U(\mathfrak{g}) \otimes \mathcal{H}^2$ by $U(\mathfrak{g}) \otimes (\mathcal{H}^2 \oplus \mathcal{H}^2 \gamma_5)$. Actually in axially extended non-Abelian MMG theories ghost, anti-ghost, gauge and axial gauge fields all have colors.

Now according to (145) one can work out a noncommutative version for Stora-Zumino descent equations. This leads to a description for anomalous behaviors of TNG theories in the setting of geometric quantization. To do so one should initially extract the Chern characters for Cartan connection form $A \in U(\mathfrak{g}) \otimes (\mathcal{H}^2 \oplus \mathcal{H}^2 \gamma_5) \otimes \Omega^1_{\text{deR}/\mathcal{SP}}(M)$. As mentioned above, $d_m$ defines a flat connection on $\Gamma(S(M)^V_N)$. Thus, $d_m + iA$ defines a connection on $\Gamma(S(M)^V_N)$ with curvature; $\hat{R} = idA - A^2$. According to [78], up to some factors, the Chern characters for this connection are $Tr\{\hat{R}^k\}$, $k \in \mathbb{N}$, for $Tr$ a trace operator on $U(\mathfrak{g}) \otimes (\mathcal{H}^2 \oplus \mathcal{H}^2 \gamma_5)$ defined by;

$$Tr\{\varpi\} := \sum_{i,\alpha,\beta,\gamma} tr\{\prod_j \Gamma^i_{\gamma_j}\} \xi_N tr\{\prod_j a^i_{\alpha_j}\} \xi_N tr\{\prod_{\downarrow j} S(b^i_{\beta_j})\},$$

(147)

for $\varpi = \sum_i \Gamma^i \otimes \varpi^i_1 \otimes \varpi^i_2 \in U(\mathfrak{g}) \otimes (\mathcal{H}^2 \oplus \mathcal{H}^2 \gamma_5)$ with $\Gamma^i = \sum_\gamma \Gamma^i_{\gamma_1} \otimes \ldots \otimes \Gamma^i_{\gamma_{n_\gamma}} \in U(\mathfrak{g})$, $\varpi^i_1 = \sum_\alpha a^i_{\alpha_1} \otimes \ldots \otimes a^i_{\alpha_{n_\alpha}} \in \mathcal{H} \oplus \mathcal{H}\gamma_5$ and $\varpi^i_2 = \sum_\beta b^i_{\beta_1} \otimes \ldots \otimes b^i_{\beta_{n_\beta}} \in \mathcal{H}$. On the other hand, $\prod_{i=1}^k c_i$ is the ordinary notation for $c_1 \ldots c_k$ while $\prod_{\downarrow i=1}^k c_i$ is used for $c_k \ldots c_1$.

The Bianchi identity [78, 79];

$$d\hat{R} = \hat{R}A - A\hat{R},$$

(148)

for $\hat{R} = idA - A^2$, the curvature 2-form, shows that $Tr\{\hat{R}^{2n+2}\}$ is a closed deRham form. While $H^{2n+2}_{deR}(\mathbb{R}^D \times M) = 0$ asserts that; $d\Omega_{2n+1}(A) = Tr\{\hat{R}^{n+1}\}$. Moreover, according to (145) one finds;

$$\delta\hat{R} = i\hat{R}\omega - i\omega\hat{R}.$$

(149)

Note that in (149), $\omega$ is the extended colored ghost field which includes colored ghost and anti-ghost fields in itself. Similarly, (149) implies that $Tr\{\hat{R}^{2n+2}\}$ is also a BRST closed form. Thus by $H^{2n+1}_{deR}(\mathbb{R}^D \times M) = 0$, one finds that; $d\Omega^1_{2n}(A) = \delta\Omega_{2n+1}(A)$, according to (146). Conventionally, $\Omega^1_{2n}(A)$, can be used as the noncommutative version of consistent anomaly [100]. The next step of descent equations is also permitted since; $2 \leq 2(n - d)$. In fact, the descent equations hold for deRham-BRST forms over $\mathbb{R}^D \times M$, $D \geq 0$, up to the term $\Omega^{2d}_{2d}(A)$. Actually, since $H^{2n-k}_{deR}(\mathbb{R}^D \times M) = 0$ for $k < 2(n - d)$, any closed $2n - k$-form over $\mathbb{R}^D \times M$ is exact if $k < 2(n - d)$. On the other hand, since $H^k_{\text{deR}}(\mathbb{R}^D \times M) = \mathbb{R}^{\binom{2d}{k}}$ for $k \leq 2d$, the exactness of closed deRham forms dies out generally. Therefore, the consistent Schwinger term $\Omega^2_{2n-1}(A)$ due to the deRham-BRST equation $d\Omega^2_{2n-1}(A) = \delta\Omega^1_{2n}(A)$ is well-defined for all MMG theories with commutative time coordinate. As mentioned above, this may lead to a well-defined setting of consistent anomalies and Schwinger terms for TNG theories.

The Chern-Simons term, $\Omega_{2n+1}(A)$, the consistent anomaly, $\Omega^1_{2n}(A)$, and the consistent Schwinger term, $\Omega^2_{2n-1}(A)$, extracted from the noncommutative version of descent equations are defined by analogy with ordinary gauge theories. In the next subsection, it is shown that the quantization of MMG theories by modified partition functions, confirms these definitions.

### 5.2. Modified Partition Functions and Consistent Anomalies of Matrix Modeled Gauge Theories

In this subsection, it is shown that the idea of modified partition functions enables one to work out the standard form of consistent anomalies and Schwinger terms for TNG theories in the setting of matrix modeling formulation. Indeed, the Chern-Simons form, ghost/anti-ghost anomalies and all types of consistent Schwinger terms can be extracted for MMG theories from the modified partition functions in complete agreement with the results of geometric quantization. In the other words, it is shown that the quantization due to modified partition functions coincides with geometric quantization in anomalous behaviors.

As a convention from now on up to the end of this subsection, $A$ is used for axially extended gauge fields. Moreover, $\omega$ is used similarly for ghost and anti-ghost fields simultaneously. Indeed $A$ and $\omega$ both have color and axial indices.

The quantization of MMG theories may naturally lead to a set of anomalous behaviors in classical symmetries. According to [102], to quantize a MMG theory and eventually to extract the anomalous terms for a MMG theory, one needs to write down its modified partition function. On the other hand, to find the modified partition function for a MMG theory, it should be used the idea of DGA with partial integration structure (DGAPI) [103]. Essentially a DGAPI is a quintuple $(\tilde{\Xi}, d, \int., \tilde{\Xi}^*, \partial)$ such that;

- a) $(\tilde{\Xi}, d)$ is a DGA.
- b) $\tilde{\Xi}^* = \oplus_{k=0}^{\infty} \tilde{\Xi}^{*-k}$ is a graded cycle, i.e. an $\mathbb{N}_0$-graded vector space with a linear map $\partial: \tilde{\Xi}^* \to \tilde{\Xi}^*$ such that $\partial Q^{-k} \in \tilde{\Xi}^{*-k-1}$ for all $Q^{-k} \in \tilde{\Xi}^{*-k}$ and $\partial^2 = 0$.
- c) There is a bilinear map $\int: \tilde{\Xi}^* \times \tilde{\Xi} \to \mathbb{C}$, $(Q, \varpi) \to \int_Q \varpi$ such that;

$$\int_Q d\varpi = \int_{\partial Q} \varpi \quad , \quad \int_Q \varpi_k \varpi_l = (-)^{kl} \int_Q \varpi_l \varpi_k$$

(150)

for all $\varpi \in \tilde{\Xi}$, differential $k$-form (resp. $l$-form) $\varpi_k$ (resp. $\varpi_l$) and $Q \in \tilde{\Xi}^*$.

Choose a large enough $D \geq 2$, for extra dimensions similar to [102] and set $\tilde{\Xi} = \oplus_{k=0} \tilde{\Xi}^k$ with $\tilde{\Xi}^k := U(\mathfrak{g}) \otimes (\mathcal{H}^2 \oplus \mathcal{H}^2 \gamma_5) \otimes \Lambda^k \mathcal{T}(\mathbb{R}^D \times M)$. Next consider $d = d_m$ as the exterior derivation operator. Also let $\tilde{\Xi}^{*-k}$ to be the real 2-dimensional vector space generated with two elements $e_0^k$ ($\partial e_0^k = 0$) and $e_1^k$ ($\partial e_1^k = e_0^{k+1}$). Then for $\varpi = \sum_i \varpi^i dx^{i_1} \wedge ... \wedge dx^{i_p}$, $\varpi^i \in U(\mathfrak{g}) \otimes (\mathcal{H}^2 \oplus \mathcal{H}^2 \gamma_5)$ define;

$$\int_{e_l^k} \varpi$$

$$:= \sum_i \int_{\mathbb{R}_l \times \mathbb{R}^{D+2(n-d)-k-1}} Tr\{\varpi^i\} \, i_k^*(dx^{i_1} \wedge ... \wedge dx^{i_p}{}_{/\mathbb{T}^{2d}}) ,$$

(151)

where;

$$dx^{i_1} \wedge \ldots \wedge dx^{i_p}{}_{/\mathbb{T}^{2d}} := \begin{cases} dx^{i'_1} \wedge \ldots \wedge dx^{i'_{p-2d}}, & dx^{i_1} \wedge \ldots \wedge dx^{i_p} = dx^{i'_1} \wedge \ldots \wedge dx^{i'_{p-2d}} \wedge \omega_{\mathbb{T}^{2d}} \\ 0, & \text{otherwise} \end{cases},$$

(152)

for $\omega_{\mathbb{T}^{2d}}$ the Riemannian volume form of $\mathbb{T}^{2d}$. Also in (151) $l$ takes the values of 0 and 1, with $\mathbb{R}_0 = \mathbb{R}$ and $\mathbb{R}_1 = \mathbb{R}_{\geq 0}$ and $i_k : \mathbb{R}^{D+2(n-d)-k} \hookrightarrow \mathbb{R}^{D+2(n-d)}$ is the inclusion map with; $(a_1, \ldots, a_{D+2(n-d)-k}) \mapsto (0, \ldots, 0, a_1, \ldots, a_{D+2(n-d)-k})$, for $k$ zeros. It can be checked that (150) holds by (151) and thus $(\tilde{\Xi}, d, \int ., \tilde{\Xi}^*, \partial)$ provides a well-defined DGAPI. Indeed the integration formula (151) defines a $D + 2(n-d)$-dimensional partial integration over $(\tilde{\Xi}, d)$ (i.e. the support of $\int$ is in $\oplus_{k=0}^{D+2(n-d)} \tilde{\Xi}^{*-k} \times \tilde{\Xi}^{D+2(n-d)-k}$).

To work out the modified partition function for a given MMG theory one should initially write down its standard partition function for a given back ground field $A \in$ Hermitian elements of $\tilde{\Xi}^1$. Indeed, since $\mathfrak{m}_N$ includes no derivations with respect to time, then according to [104], the standard partition function is given by the ordinary following form;

$$Z(A) = \int D\psi D\psi^\dagger e^{iS_{Matter}(A)}$$

(153)

Thus as was stated above the simplest case is;

$$Z(A) = \int D\psi D\psi^\dagger e^{i \int_{\mathbb{R}^{2(n-d)}} \xi_N tr\{\overline{[\psi]}_N i\gamma^\mu [\partial_\mu \psi]_N - \overline{[\psi]}_N \gamma^\mu A_\mu [\psi]_N\}},$$

(154)

where $A_\mu \in U(\mathfrak{g}) \otimes (\mathcal{H}^2 \oplus \mathcal{H}^2 \gamma_5)$. Note that the Berezin path integral is taken over the set of all possible spinors $\psi$ and their conjugate momenta $\psi^\dagger$ and not over the entries of $[\psi]_N$ and $[\psi]_N^\dagger$.

Now it is the time to modify (154) with the four axioms of [102]; a) the axiom of Coherency, b) the axiom of Relativity, c) the axiom of Gauge Invariance and finally d) the axiom of Flatness. According to [102], to apply the first axiom, the Frechet derivative should be modified with;

$$\frac{\delta}{\delta_P f(x,y)}[f]_N := \delta_{\dim P}^{2n+q}[\delta^{(2n+q)}(x,y)]_N,$$

(155)

for $f \in \mathcal{S}(\mathbb{R}^{2(n-d)+q} \times \mathbb{T}^{2d})$, and $P \subseteq \mathbb{R}^{2(n-d)+q} \times \mathbb{T}^{2d}$ an embedded closed submanifold. It can be easily seen that the given four axioms together with (155) produce a set of sufficient conditions to modify the partition function (154) similar to [102]. The Bianchi identity (148) asserts that any linear combination of elements $\prod_{i=1}^{m} Tr\{\hat{R}^{k_i}\}$, say $\Omega(A)$, is closed. Therefore, if one can show that $\Omega(A)$ is also exact then the partial integration structure reduces the integrating space due to (150). Indeed, following [102] the modified form of (154) takes the form of

$$Z_M(A) = Z(A) e^{-i \int_{e_1^{D-1}} \eta_{n,d}^N \Omega_{2n+1}(A)},$$

(156)

for $\Omega_{2n+1}(A) = d\Omega_{2n+2}(A)$, while $\Omega_{2n+2}(A)$ can be any linear combination of elements $\prod_{i=1}^{m} Tr\{\hat{R}^{k_i}\}$, $\sum_{i=1}^{m} k_i = n+1$. Also $\eta_{n,d}^N$ is a constant factor which probably depends on the matrix modeling. Here for simplicity and to keep the analogy with [102] one may suppose that $\Omega_{2n+2}(A)$ is proportional to $Tr\{\hat{R}^{n+1}\}$, the $(n+1)$th Chern character. Thus, the explicit form of consistent anomalies and Schwinger terms can be extracted for MMG theories similar to [102].

To work out the consistent anomalies for MMG theories consider an infinitesimal gauge transformation, say $\alpha$, and set $\frac{d}{dt}\big|_{t=0} Z_M(A \cdot e^{t\alpha}) = 0$, according to the gauge invariance axiom. This leads to;

$$\int_{\mathbb{R}^{2(n-d)}} \xi_N tr\{\overline{[\psi]}_N \gamma^\mu \, d_{\partial_\mu} \alpha \triangleright_* [\psi]_N\} - \eta_{n,d}^N \Omega_{2n}^1(A)(\alpha) = 0,$$

(157)

for $d\Omega^1_{2n}(A) = \delta\Omega_{2n+1}(A)$. Here the notation of $\triangleright_*$ is used for all types of actions (7)-(9). It is seen that (157) gives hand the consistent anomaly in the setting of descent equations in complete agreement with the results of the subsection 5.1.

Note that to extract the explicit form of consistent anomaly for TNG theories one should take functional derivative of (157) with respect to $\alpha(x)$, $x \in M$. This leads to emerging a Dirac delta function in the integrand which is multiplied with a functional of gauge field $A$ with star product $\star$. This may not lead to explicit form of consistent anomalies unless $\star$ be a harmonic translation-invariant product. Indeed, if $\star$ is harmonic, then according to (82), the Dirac delta function can be released from derivations inside $\star$ and then the nonintegrated consistent anomaly can be extracted precisely. But according to subsection 4.2, in the case of non-harmonic $\star$, one can replace $\star$ by its $\alpha$-cohomologous harmonic translation-invariant product $\star_H$ without losing the information. This consequently leads to well-defined form of consistent anomaly for any arbitrary TNG theory.

For the case of consistent Schwinger terms one should go further and use two infinitesimal gauge transformations $\alpha$ and $\beta$. Following [102] one can show that the consistent Schwinger terms also are given in terms of $\Omega^2_{2n-1}(A)$, as the solution of the deRham-BRST equation $d\Omega^2_{2n-1}(A) = \delta\Omega^1_{2n}(A)$. In fact, according to [102], to extract the consistent Schwinger term for a MMG theory, one should act the differential operator

$$\Xi^{(t_0;x,y)}_{2n}(\varepsilon_1, \varepsilon_2)$$
$$= i\frac{\delta}{\delta_{2n}A^a_0(t_0+\varepsilon_1,x)} i\frac{\delta}{\delta_{2n}A^b_0(t_0-\varepsilon_2,y)} - i\frac{\delta}{\delta_{2n}A^a_0(t_0-\varepsilon_2,x)} i\frac{\delta}{\delta_{2n}A^b_0(t_0+\varepsilon_1,y)}$$

(158)

on its modified partition function. This may lead to emerging one or two Dirac delta functions in the integrands which are engaged by derivations inside the star product. Then, here one can similarly use the $\alpha$-cohomologous harmonic translation-invariant product without losing the information. This will release at least one of the Dirac delta functions from derivations of star product and thus this results in well-defined form of Schwinger terms for TNG theories.

## 6. Summery and Conclusions

In this article, an elaborated formalism of gauge theories was worked out in the setting of matrix modeling. Then it was shown that these matrix modeled gauge (MMG) theories produce to an equivalent description of translation-invariant noncommutative gauge (TNG) theories. In fact, it was proven that any TNG theory can be considered as the limit of a compatible family of MMG theories. Moreover, translation-invariant star product was discussed in the setting of $\alpha$-cohomology and an algebraic version of Hodge theorem was derived for $\alpha$-cohomology groups which led to definition of harmonic translation-invariant products as the unique representing elements of $\alpha$-cohomology classes. It was also shown that loop calculations in TNG theories are entirely described by the $\alpha$-cohomology class of the translation-invariant star products. In fact, it was seen that moving through an $\alpha$-cohomology class produces no new physics. This showed that the harmonic translation-invariant products due to the Hodge theorem, play the crucial role in studding the physics of TNG theories. Then MMG theories was quantized in both geometric and path integral formulations. The geometric quantization of MMG theories led to a noncommutative version of (anti-) BRST transformations for TNG theories in the limit. It was also shown that this noncommutative counterpart of quantized symmetries, results in a concrete description of descent equations and consequently consistent anomalies and consistent Schwinger terms for TNG theories. The results of geometric quantization were confirmed by path integral quantization of modified partition functions. The explicit form of consistent anomalies and consistent Schwinger terms for a TNG theory were extracted by taking the limit of those of a compatible family of MMG theories.

# 7. Appendices

## 7.1. Appendix A

In this appendix, it is shown that the definition (59) makes sense. Moreover, (61) is proven and it is shown that; $\star_N \to \star$ as $N \to \infty$.

Consider the star product (43) over $\mathbb{R}^l$. By $\alpha(p+q,p) = \langle \alpha_L(ip)|\alpha_R(iq)\rangle$, (43) takes the form of;

$$f \star g(x) = \int \frac{d^l p}{(2\pi)^l} \frac{d^l q}{(2\pi)^l} \tilde{f}(p)\tilde{g}(q) e^{\langle \alpha_L(ip)|\alpha_R(iq)\rangle} e^{i(p+q).x}.$$

(A1)

Then, using the definition of;

$$\tilde{f}(p) = \int_{\mathbb{R}^l} f(x) e^{-ip.x},$$

(A2)

one finds that;

$$f \star g(x) = \int \frac{d^l p}{(2\pi)^l} \frac{d^l q}{(2\pi)^l} d^l y d^l z \; f(y) g(z) \; e^{\langle \alpha_L(ip)|\alpha_R(iq)\rangle} e^{ip.(x-y)} e^{iq.(x-z)}$$

$$= \int \frac{d^l p}{(2\pi)^l} \frac{d^l q}{(2\pi)^l} d^l y d^l z \; f(y) g(z) \; e^{\langle \alpha_L(-\vec{\partial}^y)|\alpha_R(-\vec{\partial}^z)\rangle} e^{ip.(x-y)} e^{iq.(x-z)}$$

$$= \int \frac{d^l p}{(2\pi)^l} \frac{d^l q}{(2\pi)^l} d^l y d^l z \; e^{ip.(x-y)} e^{iq.(x-z)} \; e^{\langle \alpha_L(\vec{\partial}^y)|\alpha_R(\vec{\partial}^z)\rangle} f(y) g(z)$$

$$= \int d^l y d^l z \; \delta^{(m)}(x-y) \delta^{(m)}(x-y) \; e^{\langle \alpha_L(\vec{\partial}^y)|\alpha_R(\vec{\partial}^z)\rangle} f(y) g(z).$$

(A3)

Therefore, it is shown that;

$$f \star g(x) = \pi \left( \sum_{p=0}^{\infty} \frac{1}{p!} \left( \sum_{i=1}^{\infty} \alpha_L \left( \vec{\partial}|_x \right)^i \otimes \alpha_R \left( \vec{\partial}|_x \right)^i \right)^p (f \otimes g) \right),$$

(A4)

with good agreement with (61). Obviously, when $|\alpha_L\rangle$ and $|\alpha_R\rangle$ are finite dimensional vectors, the upper limit of the inner summation is a finite number. Note that (A4) induces a quantization structure on $C^\infty(\mathbb{T}^l)$ as a subset of $C^\infty(\mathbb{R}^l)$ made up of smooth periodic functions on $\mathbb{R}^l$. Thus, from now on for simplicity we may focus most of our attention to $C^\infty(\mathbb{T}^l)$. As mentioned above we also restrict ourselves to Schwartz functions $\mathcal{S}(\mathbb{T}^l)$.

To prove (A4) for translation-invariant noncommutative structures over $\mathbb{T}^l$, it is enough to note that;

$$\sum_{\vec{p}} e^{ik_{\vec{p}} \cdot (x-y)} = (2\pi R)^l \delta^{(l)}(x-y)$$

(A5)

for $x, y \in \mathbb{T}^l$ and also for $k_{\vec{p}}$ and $\vec{p}$ as defined in (57). This with (58) and (A3) prove (A4) for translation-invariant noncommutative structures over $\mathbb{T}^l$.

Now it is the time to prove (61). To this end, note that for any two Schwartz functions $f, g \in \mathcal{S}(\mathbb{T}^l)$, for any Fourier mode $k_{\vec{p}}$, and finally for any differential operator $\Xi$, (A4) and (A5) imply that;

$$\int_{x \in \mathbb{T}^l} e^{-ik_{\vec{p}} \cdot x} (f \star \Xi(g))(x)$$

$$= \frac{1}{(2\pi R)^l} \sum_{(\mu), \vec{q}} \frac{1}{|\mu|!} \int_{x \in \mathbb{T}^l} e^{-i(k_{\vec{p}} - k_{\vec{q}}) \cdot x} \alpha_L \left( \vec{\partial}|_x \right)^{(\mu)} (f) \int_{y \in \mathbb{T}^l} e^{-ik_{\vec{q}} \cdot y} \alpha_R \left( \vec{\partial}|_y \right)^{(\mu)} (\Xi(g)).$$

(A6)

To prove (61), at the first step one has to show that the definition (59) makes sense. It is claimed that for any set of Schwartz functions, say $\{f_i\}_{i=1}^{k \geq 2}$;

$$\sum_{(\sigma_i), \vec{r}_i, i=1,\ldots,k} [f_1]_{N\vec{p}\vec{r}_1}^{(\mu)(\sigma_1)} [f_2]_{N\vec{r}_1\vec{r}_2}^{(\sigma_1)(\sigma_2)} \cdots [f_k]_{N\vec{r}_{k-1}\vec{q}}^{(\sigma_{k-1})(\nu)}$$

$$= \zeta^{(\mu)(\nu)} \int_{x \in \mathbb{T}^l} e^{-ik_{\vec{p}} \cdot x} \alpha_R{}^{(m)} \left(\vec{\partial}\big|_x\right)^{(\mu)} \left(f_1 \star \cdots \star \left(e^{ik_{\vec{q}} \cdot x} \alpha_L{}^{(m)} \left(\vec{\partial}\big|_x\right)^{(\nu)} (f_k)\right)\right)$$

$$+ \cdots$$

(A7)

for $\zeta^{(\mu)(\nu)} = \frac{1}{\sqrt{|\mu|!|\nu|!}(2\pi R)^l}$ and for the dotted term which vanishes as $N \to \infty$.

To prove (A7) one needs the induction. By (A6) and the revised form of (57);

$$[f]_{N\vec{p}\vec{q}}^{(\mu)(\nu)}$$

$$= \frac{1}{\sqrt{|\mu|!|\nu|!}(2\pi R)^l} \int_{x \in \mathbb{T}^l} e^{-ik_{\vec{p}} \cdot x} \alpha_R{}^{(m)} \left(\vec{\partial}\big|_x\right)^{(\mu)} \left(e^{ik_{\vec{q}} \cdot x} \alpha_L{}^{(m)} \left(\vec{\partial}\big|_x\right)^{(\nu)} (f)\right)$$

$$= \frac{\alpha_R{}^{(m)}(ik_{\vec{p}})^{(\mu)}}{\sqrt{|\mu|!|\nu|!}(2\pi R)^l} \int_{x \in \mathbb{T}^l} e^{-i(k_{\vec{p}} - k_{\vec{q}}) \cdot x} \alpha_L{}^{(m)} \left(\vec{\partial}\big|_x\right)^{(\nu)} (f),$$

(A8)

for any $f \in \mathcal{S}(\mathbb{T}^l)$, one finds

$$\sum_{(\sigma), \vec{r}} [f_1]_{N\vec{p}\vec{r}}^{(\mu)(\sigma)} [f_2]_{N\vec{r}\vec{q}}^{(\sigma)(\nu)}$$

$$= \frac{\alpha_R{}^{(m)}(ik_{\vec{p}})^{(\mu)}}{\sqrt{|\mu|!|\nu|!}(2\pi R)^l} \int_{x \in \mathbb{T}^l} e^{-ik_{\vec{p}} \cdot x} f_1 \star \left(e^{ik_{\vec{q}} \cdot x} \alpha_L{}^{(m)} \left(\vec{\partial}\big|_x\right)^{(\nu)} (f_2)\right) + \cdots,$$

(A9)

for any two Schwartz functions $f_1, f_2 \in \mathcal{S}(\mathbb{T}^l)$. Then, we conclude that;

$$\sum_{(\sigma),\vec{r}} [f_1]_{N\vec{p}\vec{r}}^{(\mu)(\sigma)} [f_2]_{N\vec{r}\vec{q}}^{(\sigma)(\nu)}$$

$$= \zeta^{(\mu)(\nu)} \int_{x \in \mathbb{T}^l} e^{-ik_{\vec{p}} \cdot x} \alpha_R{}^{(m)} \left(\vec{\partial}\big|_x\right)^{(\mu)} \left( f_1 \star \left( e^{ik_{\vec{q}} \cdot x} \alpha_L{}^{(m)} \left(\vec{\partial}\big|_x\right)^{(\nu)} (f_2) \right) \right) + \cdots .$$

(A10)

Note that the dotted term in (A8) and (A9) comes from three approximations;

- a) The Dirac delta function in (A5) comes from summation over all the Fourier modes while in (A10) the summation is taken place over finitely many Fourier modes.
- b) $|\alpha_L\rangle$ and $|\alpha_R\rangle$ are approximated by $|\alpha_L{}^{(m)}\rangle$ and $|\alpha_R{}^{(m)}\rangle$ respectively.
- c) Finitely many terms of the Taylor expansion of $e^{\langle \alpha_L | \alpha_R \rangle}$ are considered in (A10).

Clearly the dotted term vanishes rapidly as $N(m) \to \infty$. However, the rest of the proof is trivial. By induction suppose (A7) has already been proved for any $1 \leq k < K$. Thus;

$$\sum_{(\sigma_i),\vec{r}_i, i=1,\ldots,K} [f_1]_{N\vec{p}\vec{r}_1}^{(\mu)(\sigma_1)} [f_2]_{N\vec{r}_1\vec{r}_2}^{(\sigma_1)(\sigma_2)} \cdots [f_K]_{N\vec{r}_{K-1}\vec{q}}^{(\sigma_{K-1})(\nu)}$$

$$= \sum_{\vec{r}_1,(\sigma_1)} \frac{\alpha_R{}^{(m)}(ik_{\vec{p}})^{(\mu)}}{\sqrt{|\mu|!\,|\sigma_1|!}\,(2\pi R)^l} \eta^{(\sigma_1)(\nu)} \int_{x \in \mathbb{T}^l} e^{-i(k_{\vec{p}} - k_{\vec{r}_1}) \cdot x} \alpha_L \left(\vec{\partial}\big|_x\right)^{(\sigma_1)} (f_1) \times$$

$$\left( \int_{y \in \mathbb{T}^l} e^{-ik_{\vec{r}_1} \cdot y} \alpha_R{}^{(m)} \left(\vec{\partial}\big|_y\right)^{(\sigma_1)} \left( f_2 \star \ldots \star \left( e^{ik_{\vec{q}} \cdot y} \alpha_L{}^{(m)} \left(\vec{\partial}\big|_y\right)^{(\nu)} (f_K) \right) \right) + \cdots \right).$$

(A11)

Using (A6) one concludes that;

$$\sum_{(\sigma_i),\vec{r}_i,i=1,\ldots,K} [f_1]_{N\vec{p}\vec{r}_1}^{(\mu)(\sigma_1)} [f_2]_{N\vec{r}_1\vec{r}_2}^{(\sigma_1)(\sigma_2)} \ldots [f_K]_{N\vec{r}_{K-1}\vec{q}}^{(\sigma_{K-1})(\nu)}$$

$$= \zeta^{(\mu)(\nu)} \int_{x \in \mathbb{T}^l} e^{-ik_{\vec{p}}\cdot x} \alpha_R^{(m)} \left(\vec{\partial}\big|_x\right)^{(\mu)} \left( f_1 \star \ldots \star \left( e^{ik_{\vec{q}}\cdot x} \alpha_L^{(m)} \left(\vec{\partial}\big|_x\right)^{(\nu)} (f_K) \right) \right)$$

$$+ \cdots .$$

(A12)

Therefore (A7) follows. Finally taking the trace of (A52) one finds out that;

$$tr\{[f_1]_N \ldots [f_K]_N\}$$

$$= \sum_{\vec{p},(\mu)} \zeta^{(\mu)(\mu)} \alpha_R^{(m)} (ik_{\vec{p}})^{(\mu)} \int_{x \in \mathbb{T}^l} e^{-ik_{\vec{p}}\cdot x} f_1 \star \ldots \star \left( e^{ik_{\vec{p}}\cdot x} \alpha_L^{(m)} \left(\vec{\partial}\big|_x\right)^{(\mu)} (f_K) \right) + \cdots$$

$$= \sum_{\vec{p},(\mu)} \frac{\alpha_R^{(m)} (ik_{\vec{p}})^{(\mu)}}{|\mu|! \, (2\pi R)^l} \int_{x \in \mathbb{T}^l} e^{-ik_{\vec{p}}\cdot x} f_1 \star \ldots \star \left( e^{ik_{\vec{p}}\cdot x} \alpha_L^{(m)} \left(\vec{\partial}\big|_x\right)^{(\mu)} (f_K) \right) + \cdots .$$

(A13)

As $R \to \infty$ then $k_{\vec{p}} \to (0, \ldots, 0)$ for any $\vec{p}$, while $|\alpha_R^{(m)}\rangle$ vanishes at the origin. Therefore, the summation over multi-plets in (A53) becomes restricted to the empty multi-plet ($\emptyset$). This leads to;

$$\lim_{R \to \infty} (2\pi R)^l \times tr\{[f_1]_N \ldots [f_K]_N\} = \sum_{\vec{p}} \int_{x \in \mathbb{T}^l} f_1 \star \ldots \star f_K + \cdots$$

$$= (2m+1)^l \int_{x \in \mathbb{T}^l} f_1 \star \ldots \star f_K + \cdots ,$$

(A14)

for $(2m + 1)^l$ numbers of Fourier modes included in the summation. This proves that the definition (59) makes sense and thus (61) follows naturally.

## 7.2. Appendix B

In this appendix, it is shown that $\alpha \in C^2(\mathbb{R}^m)$ is a coboundary if;

$$\partial_2 \alpha = \partial \alpha = 0 ,$$

$$\alpha(p, q) = \alpha(p, p - q) ,$$

$$\alpha(p, q) = -\alpha(-p, -q) ,$$

$$\alpha(p, q) = -\alpha(q, p) ,$$

(B1)

for any $p, q \in \mathbb{R}^m$. To this end initially, choose the coordinate system of;

$$\begin{cases} w = p - q \\ z = q \end{cases}.$$

(B2)

It can be shown that;

$$\frac{\partial^2 \alpha}{\partial z^i \partial w^j}(p, q) = \frac{\partial^2 \alpha}{\partial q^i \partial p^j}(p, 0)$$

(B3)

for any $p, q \in \mathbb{R}^m$ and for any $1 \leq i, j \leq m$. Indeed

$$\frac{\partial^2 \alpha}{\partial z^i \partial w^j}(p, q) = \frac{\partial^2 \alpha}{\partial p^i \partial p^j}(p, q) + \frac{\partial^2 \alpha}{\partial q^i \partial p^j}(p, q)$$

$$= \frac{d^2}{drds}\bigg|_{r=s=0} \alpha(p + re^j + se^i, q + se^i).$$

(B4)

for $\{e^i\}_{i=1}^m$, the standard basis of $\mathbb{R}^m$. Using (44) one finds that;

$$\frac{d^2}{drds}\bigg|_{r=s=0} \alpha(p+re^j+se^i, q+se^i)$$

$$= \frac{d^2}{drds}\bigg|_{r=s=0} \left( \alpha(p+re^j, q) - \alpha(p+re^j, -se^i) - \alpha(q, -se^i) \right)$$

$$= \frac{\partial^2 \alpha}{\partial q^i \partial p^j}(p, 0) \,.$$

(B5)

Therefore;

$$\frac{\partial}{\partial q^k} \frac{\partial^2}{\partial z^i \partial w^j} \alpha = 0$$

(B6)

for any $1 \leq i, j, k \leq m$. In the other words, one finds that;

$$\left( \frac{\partial}{\partial z^k} - \frac{\partial}{\partial w^k} \right) \frac{\partial^2}{\partial z^i \partial w^j} \alpha = 0 \,.$$

(B7)

Now suppose that;

$$\alpha(w, z) = f_1(w) + f_2(w, z) + f_3(z)$$

(B8)

for $f_1, f_2, f_3 \in C^\infty(\mathbb{R}^m)$ such that;

$$f_2(w, z) \neq g_1(w) + g_2(w, z) \,,$$

$$f_2(w, z) \neq h_1(z) + h_2(w, z) \,,$$

(B9)

$g_1, g_2, h_1, h_2 \in C^\infty(\mathbb{R}^m)$, for non-constant $g_1$ and $h_1$. Thus (B7) leads to;

$$\left(\frac{\partial}{\partial z^k} - \frac{\partial}{\partial w^k}\right)\frac{\partial^2}{\partial z^i \partial w^j} f_2 = 0 .$$

(B10)

Now choose the coordinate system of

$$\begin{cases} \eta = z + w \\ \xi = z - w \end{cases}.$$

(B11)

Therefore, by (B10) and (B11) one finds that;

$$\frac{\partial}{\partial \xi^k} \frac{\partial^2}{\partial z^i \partial w^j} f_2 = 0 .$$

(B12)

Consequently, we have;

$$\frac{\partial^2 f_2}{\partial z^i \partial w^j}(\eta, \xi) = h_{ij}(\eta) .$$

(B13)

Then since

$$\frac{\partial^2}{\partial z^i \partial w^j} = \frac{\partial^2}{\partial \eta^i \partial \eta^j} - \frac{\partial^2}{\partial \eta^i \partial \xi^j} + \frac{\partial^2}{\partial \xi^i \partial \eta^j} - \frac{\partial^2}{\partial \xi^i \partial \xi^j},$$

(B14)

(B13) lets one to set; $f_2(\eta, \xi) = g(\eta) + \xi^i g_i(\eta) + \xi^i \xi^j g_{ij}(\eta)$, $1 \leq i, j \leq m$, for some $g, g_i, g_{ij} \in C^\infty(\mathbb{R}^m)$. Therefore;

$$\alpha(w, z) = f_1(w) + g(z + w) + (z - w)^i g_i(z + w)$$
$$+ (z - w)^i (z - w)^j g_{ij}(z + w) + f_3(z) .$$

(B15)

Note that since $\alpha$ is commutative then; $\alpha(w,z) = \alpha(z,w)$. This implies that;

$$(f_3(w) - f_1(w)) - (f_3(z) - f_1(z)) = 2(z-w)^i g_i(z+w),$$

(B16)

Then, (B9) leads to;

$$g_i = 0, \quad f_1 = f_3 = f.$$

(B17)

Therefore;

$$\alpha(p,q) = f(q) + g(p) + (2q-p)^i(2q-p)^i g_{ij}(p) + f(p-q),$$

(B18)

for any $p, q \in \mathbb{R}^m$. Now by $\alpha(-p,-q) = \alpha(q,p)$ one finds that;

$$f(-q) + g(-p) + (2q-p)^i(2q-p)^i g_{ij}(-p)$$
$$= f(p) + g(q) + (2p-q)^i(2p-q)^i g_{ij}(q).$$

(B19)

Acting $\dfrac{\partial^2}{\partial p^i \partial q^j}$ on both sides of (B19) yields the following result

$$4g_{ij}(-p) + 4(2q-p)^k \frac{\partial g_{jk}}{\partial p^i}(-p) = 4g_{ij}(q) - 4(2p-q)^k \frac{\partial g_{ik}}{\partial q^j}(q).$$

(B20)

Thus, one finds that;

$$\frac{\partial g_{jk}}{\partial p^i}(p) = -\frac{\partial g_{ik}}{\partial p^j}(p).$$

(B21)

Acting $\dfrac{\partial}{\partial q^k}$ on both sides of (B19) and using (B21) one also finds that;

$$\frac{\partial g_{jk}}{\partial p^i}(-p) = \frac{1}{2}(2p-q)^l \frac{\partial^2 g_{il}}{\partial q^k \partial q^j}(q).$$

(B22)

Setting $q = 2p$, gives rise to;

$$g_{ij}(p) = c_{ij} \in \mathbb{C}.$$

(B23)

Thus

$$f_2(w, z) = g(w + z) - 2w^i z^j c_{ij} + w^i w^j c_{ij} + z^i z^j c_{ij},$$

(B24)

which according to (B9) yields $c_{ij} = 0$ for any $i$ and $j$. Therefore;

$$f_2(w, z) = g(w + z).$$

(B25)

Consequently, by (B19) it follows that;

$$f(-q) - g(q) = f(p) - g(-p),$$

(B26)

which results in;

$$g(p) = f(-p) + c_0,$$

(B27)

for any $p \in \mathbb{R}^m$ and for a complex number $c_0$. Thus according to (B18) we have;

$$\alpha(p, q) = f(q) + f(-p) + f(p - q) + c_0$$

(B28)

for any $p, q \in \mathbb{R}^m$. But (45) leads to;

$$f(-p) = -f(p) + c_1,$$

(B29)

for any $p \in \mathbb{R}^m$ and for $c_1 = -f(0) - c_0$. Thus, (B28) takes the following form;

$$\alpha(p,q) = f(q) - f(p) + f(p-q) + c_1.$$

(B30)

Finally the equality $\alpha(p,q) = -\alpha(-p,-q)$ implies that $c_1 = 0$ and eventually

$$\alpha(p,q) = f(q) - f(p) + f(p-q),$$

(B31)

for any $p, q \in \mathbb{R}^m$. On the other hand, by (45) we have; $f(0) = 0$. Therefore, $f \in C^1(\mathbb{R}^m)$ and consequently; $\alpha = \partial f$.

## 7.3. Appendix C

In this appendix, it is shown that (120) concretely leads to (123). To see this note that by (120);

$$\omega(p,p) = \omega(np,p) = 0,$$

(C1)

for any $n \in \mathbb{Z}$. Then, the iterated form of (C1) would be

$$\omega(N_k p, N_{k-1} p) = 0,$$

(C2)

$k \geq 1$, for recursive formulae;

$$N_k = n_k N_{k-1} + N_{k-2},$$

(C3)

with $n_k \in \mathbb{Z}$, $k \geq 2$, $N_0 = 1$ and $N_1 = n_1 \in \mathbb{Z}$. Clearly (C1) can be rewritten in the form of

$$\omega\left(\frac{N_k}{N_{k-1}}p, p\right) = 0.$$

(C4)

But according to (C3) it can be seen that;

$$\frac{N_k}{N_{k-1}} = n_k + \cfrac{1}{n_{k-1} + \cfrac{1}{n_{k-2} + \cfrac{1}{\ddots \, n_2 + \cfrac{1}{n_1}}}}.$$

(C5)

On the other hand, it is known [105] that for any rational number $r \in \mathbb{Q}$, there is a finite sequence of integers $n_i$, $1 \leq i \leq k$, with $n_k \in \mathbb{Z}$ and $n_i > 0$ for $1 \leq i < k$, such that

$$r = \frac{N_k}{N_{k-1}}$$

(C6)

in accordance to (C5). This together with (C4) proves (123).

As mentioned in subsection 4.2, the continuity of $\omega$ consequently leads to (124). On the other hand, (124) shows that any translation-invariant product on $C^\infty(\mathbb{R})$, is $\alpha$-cohomologous to ordinary point-wise product and therefor is commutative. In the other words (124) results in $H^2_\alpha(\mathbb{R}) = 0$, which asserts that there are no translation-invariant noncommutative star products on $C^\infty(\mathbb{R})$, which is not obvious at all.

# Acknowledgement

The author should say his gratitude to F. Ardalan, L. Bonora, A. Connes, H. Grosse, M. Hayakawa, M. Khalkhali, T. Krajewski, E. Langmann, C. Martin, J. Maldacena, S. Minwalla, V. Rivasseau, N. Seiberg, R. J. Szabo, A. Tanasa, B. Tsygan, F. Vignes-Tourneret, P. Vitale, A. Weinstein, E. Witten and M. Wohlgenannt for hints and discussions. Also this article owes most of its appearance to S. Ziaee for many things. Finally my special thanks go to Ahmad Shafiei Deh Abad for his ever warm welcom, patience and useful comments in geometry.